\documentclass{tlp}
\usepackage{aopmath}
\usepackage{epsf}

\newcommand{\Staerk}{St\"{a}rk}

\newcommand{\And}{ \& } 
\newcommand{\Or}{\vee}
\newcommand{\Not}{\neg}

\newtheorem{thm}[theorem]{Theorem}
\newtheorem{cor}[theorem]{Corollary}
\newtheorem{prop}[theorem]{Proposition}

\newcommand{\Sep}{\hspace*{5mm}}
\newcommand{\Indent}{\hspace*{5mm}}

\newsavebox{\colondashbox}
\savebox{\colondashbox}{{\tt \ :-\ }}
\newcommand{\colondash}{\usebox{\colondashbox}}

\newcommand{\Rw}[1]{\rhd_{#1}}
\newcommand{\dfnf}{{\it dfnf}}

\newenvironment{recdef}%
{\begin{list}{$\bullet$}{%
\setlength{\topsep}{0.5mm}%
\setlength{\itemsep}{0pt}%
\setlength{\parsep}{0.5mm}}%
}%
{\end{list}}

\newcommand{\Goes}[1]{~\Rightarrow_{#1}~}

\newlength{\capwidth}
\newlength{\restwidth}
\newcommand{\Cap}[1]{\makebox[\capwidth][l]{#1}}
\newcommand{\Rule}[4]{{\samepage%
\Cap{#1}%
\parbox{\restwidth}{\[\frac{#2}{#3}\]}%
\\ \Cap{~}\parbox{\restwidth}{#4}}}

\begin{document}
\bibliographystyle{tlp}

\title{The Witness Properties and the Semantics of the Prolog Cut}
\author[James H.\ Andrews]
{JAMES H.\ ANDREWS\\
Department of Computer Science,
University of Western Ontario \\
London, Ontario, Canada N6A 5B7}
\pagerange{\pageref{firstpage}--\pageref{lastpage}}
\volume{\textbf{10} (3):}
\jdate{March 2000}
\setcounter{page}{1}
\pubyear{2000}

\maketitle
\label{firstpage}

\begin{abstract}
The semantics of the Prolog ``cut'' construct is explored in the
context of some desirable properties of logic programming
systems, referred to as the witness properties.  The witness
properties concern the operational consistency of responses to
queries.  A generalization of Prolog with negation as failure
and cut is described, and shown not to have the witness properties.
A restriction of the system is then described, which preserves
the choice and first-solution behaviour of cut but allows the
system to have the witness properties.

The notion of cut in the restricted system is more restricted
than the Prolog hard cut, but retains the useful first-solution
behaviour of hard cut, not retained by other proposed cuts such
as the ``soft cut''.  It is argued that the restricted system
achieves a good compromise between the power and utility of the
Prolog cut and the need for internal consistency in logic
programming systems.  The restricted system is given an abstract
semantics, which depends on the witness properties; this
semantics suggests that the restricted system has a deeper
connection to logic than simply permitting some computations
which are logical.

Parts of this paper appeared previously in a different form in
the Proceedings of the 1995 International Logic Programming
Symposium \cite{andrews-cut-ilps95}.
\end{abstract}

\section{Introduction}

Since the first widely-used Prolog implementations of the
early 1980s, Prolog programmers have had access to some
powerful constructs for controlling the backtracking behaviour
of their programs.  The best-known of these is the
``cut'', written ``!'', which appears as a literal in the
sequence of literals in a clause body.
Cut allows programmers to direct
the flow of control in a program by cutting away backtrack
points which lead to unwanted execution paths.

Programmers have embraced cut enthusiastically.  Most large
Prolog programs now in use contain cuts, or related constructs
such as the if-then-else construct ($A$ \verb/->/ $B$ \verb/;/ $C$).
Cut is used mainly for choosing between clauses.  However,
it has other important uses, such as for obtaining the first
solution to a subgoal and discarding others.

Unfortunately, the unrestricted use of cuts produces a program
which has no direct logical interpretation.
A cut does not even have an effect
restricted to the clause in which it
appears; rather, it may affect all the clauses of the predicate
which its clause is defining.  It is therefore difficult to give
a semantics to a program which uses cut, other than an operational
semantics.

It seems therefore that the use of cut, and the constructs
related to it, must be restricted in order to regain a logical
interpretation for Prolog programs.  Various approaches to this
have been proposed, including the ``soft cut'' and the mode and
determinism restrictions of the Mercury system
\cite{mercury-jlp}.  However, neither soft cut nor Mercury allow
the behaviour of cut which allows us to choose the first
solution to a subgoal and discard other solutions.  This is a
fundamental property often used by Prolog programmers, so it
would be preferable to preserve it.

Like most logic programming researchers, we believe that
Prolog's ``hard cut'' cannot be salvaged from a logical point of
view.  However, we do not believe it is necessary to retreat all
the way to soft cut.
In this paper, we show how the hard cut of Prolog can be
restricted to produce a cut, referred to as ``firm cut'', which
has important advantages over both soft and hard cut.  Firm cut
allows useful behaviours such as first-solution which are
disallowed by soft cut.  Modulo a 
run-time or compile-time mode restriction, firm cut is
operationally identical to the more widely-used hard cut, which
soft cut is not.
However, firm cut disallows the most non-logical and anti-intuitive
behaviours of hard cut, and while (like hard cut) it has no
purely logical interpretation, it still satisfies
some important consistency properties which hard cut does not.

We refer to the consistency properties which firm cut satisfies
as the ``witness properties''.  Because it satisfies these
properties, firm cut and the systems incorporating it can be
given abstract semantics based on compositional valuation functions
(functions from goals to truth values).
We demonstrate this by giving such an
abstract semantics for the system with firm cut.

Along the way, we also introduce a form of formula, the {\it if}
formula, which allows a Prolog program with cuts to be given a
``completed form'' analogous to the Clark completion of a
definite clause program.  This form of program may have
applications even when dealing with other forms of cut.

\subsection{The Witness Properties}

One of the central properties we like to prove about logic
programming systems is the equivalence between the operational
and logical semantics.  The well-known equivalence of
SLD-resolution and the least model semantics is the most obvious
example.  Such properties show that the logic programming
system in question achieves some standard of expected behaviour.

But what if the logic programming system has no logical semantics?
Is there any standard to which such a system can be held, any
middle ground between a system with a full logical semantics and
a system indistinguishable from imperative or functional
programming systems?
We believe that there is, and suggest the {\it witness
properties} as a possible standard.

The witness properties are as follows:
\begin{enumerate}
\item (Success property) If a goal formula $G$ succeeds
  (returns an answer substitution), then some ground
  instance of $G$ succeeds.
\item (Failure property) If a goal formula $G$ fails
  (terminates without returning an answer substitution),
  then all ground instances of $G$ fail.
\end{enumerate}
The witness properties accord with our intuitions about the
internal consistency of logic programming systems, and about
the nature of formulas and the search for satisfying substitutions
for them.  They therefore provide a possible standard to which
to hold logic programming systems.
Their name comes from the notion of witness for
an existentially-quantified formula:  the formula $\exists x~G$
is true if there is a witness term $t$ such that $G[x:=t]$
is true, and false otherwise.  A goal formula such as $p(x)$
can be read as asking whether $\exists x~p(x)$ is true.

In the success property, we insist on {\it ground} instances in
particular, partly because otherwise it would always be
vacuously true: $G$ is an instance of $G$, so if $G$ succeeds,
some instance of it succeeds.  We express the failure property
in terms of ground terms as well for symmetry.  Another
reason for using ground terms in the statement of the properties
is that it allows the success and failure of goals with free
variables to be characterized in terms of the simpler notion of
success and failure of ground goals.  Many variants
of these properties are possible and may be valuable for
different applications.

Note that the converses of the witness properties are not
necessarily enjoyed by logic programming systems.
The converse of the success property (if an instance of $G$
succeeds, then $G$ succeeds) is not enjoyed by any
deterministic definite clause resolution system (like Prolog)
using a search rule which selects clauses in order, as the
following example shows:
\begin{tabbing}
\Indent{} \= \kill
\>  $p(0) \colondash p(0).$ \\
\>  $p(1).$
\end{tabbing}
The goal $p(y)$ diverges even though its instance $p(1)$
succeeds.  The converse of the failure property (if all ground
instances of $G$ fail, then $G$ fails) is not enjoyed by
any deterministic definite clause resolution system, regardless
of search or selection
rule, as the following example (based on that of Clark, Andreka
and Nemeti) shows:
\begin{tabbing}
\Indent{} \= \kill
\>  $p(f(x)) \colondash p(x).$
\end{tabbing}
The goal $p(y)$ diverges even though every ground instance
of it fails.

The witness properties also have theoretical significance.
Generally, we may consider a logic programming system
to be unsatisfying from a logical point of view if it can be
given only operational semantics, as this leads us to suspect
that the operational model is a ``hack'' which is only logical
in the sense that it permits some computations which can be
viewed as logical.
Of course, every operational semantics for an LP language can be
converted to a denotational semantics if operational notions
such as unification and substitution sequence are suitably
``reified'' (i.e., represented explicitly by mathematical
constructs).  However, these semantics should not necessarily
boost our confidence that the operational model is logical, any
more than the operational semantics did.
The existence of semantics which do not reify operational
notions suggests that we are dealing with a system which has a
deeper connection to logic than simply permitting logical
computations.  Evidence from past research and the present paper
indicates that the witness properties lead to such semantics.

\subsection{This Paper}

In this paper, we show how the hard cut of Prolog, as restricted
to ``firm cut'', retains the witness properties and can be given
an abstract, non-reifying semantics.  We believe that the
resulting system is the best compromise yet found between the
power and utility of the Prolog cut and the need for internal
consistency in logic programming systems.

In section 2, we review background and related work in more
detail.  In section 3, we present the notation and syntax we
will use for logic
programs with cut and a new construct, {\it if}.  In section 4,
we present a first operational semantics for the extended
programs.  This operational semantics corresponds to Prolog,
with its permissive, non-logical view of negation and cut; thus
it is referred to as the ``liberal''
semantics.  In section 4, we also show that the {\it if}
construct allows us to derive a convenient ``completed form''
for every program, in which each predicate is defined by exactly
one clause.

In section 5, we restrict the liberal operational semantics, and
show that the restricted system has the witness properties.
The new, restricted system is referred to as the ``conservative'' semantics,
and firm cut is defined as the cut associated with it.  In section 6, we
define a non-reifying abstract semantics for the system with
firm cut, using the witness properties to prove soundness and
completeness of the conservative semantics.  Finally, in Section
7 we give some conclusions and suggestions for further research.

\section{Background and Related Work}

In this section, we introduce the background of this research
and the other research related to it.  We have grouped this
material into three sections: one concerning the cut and other
choice constructs like the if-then-else, one concerning the
semantics of depth-first Prolog and cut, and one concerning the
various different notions of termination of a logic program.

\subsection{Cut and Other Choice Constructs}

Cut was introduced in the DECsystem-10/20 Prolog of 1982,
written by David Warren, Fernando Pereira, Lawrence Byrd and
Luis Pereira.  It was recognized even at the time as a
``meta-theoretic'' control construct, which could at best be
read as making meta-level manipulations of the search tree.  Cut
was taken into the C-Prolog interpreter \cite{c-prolog}, which
became a very widely distributed early version of the language.

Cut operates by cutting away previously-encountered
alternatives.  Consider the following program:
\begin{tabbing}
\Indent{} \= \Indent{} \= \kill
\>  $p(a, y).$ \\
\>  $p(b, y) \colondash q(y), !, r(y).$ \\
\>  $p(x, y).$\vspace{2mm} \\
\>  $q(c).$ \\
\>  $q(d).$\vspace{2mm} \\
\>  $r(d).$ \\
\end{tabbing} \vspace{-1ex}
($x$ and $y$ are variables, and $a$-$e$ are constants.)  With
respect to this program, calls to the predicate $p$ exhibit the
following behaviour.
\begin{itemize}
\item Goals of the form $p(a, t)$ succeed for any term $t$.
\item Goals of the form $p(b, t)$ succeed only if $t$ is $d$, or
  if $t$ is not unifiable with either $c$ or $d$; otherwise
  they fail.  For instance:
  \begin{itemize}
  \item The goal $p(b,y)$ fails, because $y$ is unified with $c$
    by the first clause for $q$, the last clauses for $p$ and
    $q$ are cut away, and $r(c)$ fails.
  \item The goal $p(b,d)$ succeeds because $q(d)$ succeeds, only
    the last clause for $p$ is cut away, and $r(d)$ succeeds.
  \item The goal $p(b,b)$ succeeds because $q(b)$ fails entirely,
    and so the third clause for $p$ is used.
  \end{itemize}
\item Finally, goals of the form $p(s,t)$, where $s$ is anything
  other than $a$ and $b$, succeed.
\end{itemize}

Cut therefore cuts away not only the later clauses of the same
predicate, but also the alternative clauses for subgoals that
appear earlier in the clause.  The former behaviour allows us to
select clauses, but the latter behaviour allows us to choose the
first solution to a subgoal (by stating the subgoal and
following it by a cut).  This may be used for various reasons:  to
discard solutions that we, the programmers, know to be equivalent
to the first; to prevent backtracking because we know there will
be no more successes; or simply to select the first solution
because we know that is the one we are interested in
(for instance, ``$prime(x), x > 100, !$'' for the first prime
greater than $100$).

We can see immediately that Prolog with the form of cut
described above does not have the failure witness property,
since $p(b,y)$ fails but $p(b,d)$ succeeds.  (Examples can be
constructed violating the success witness property as well.)
The most common way to fix this problem with cut is to allow backtracking
into the portion before the cut -- that is, to cut away later
clauses to the current clause but not alternative clauses to
subgoals before the cut.  This is generally referred to as the
``soft cut'', and the more usual cut is referred to as the
``hard cut'' in order to distinguish it.
With soft cut, we can regain a logical interpretation:
if ! in the above program is interpreted
as soft cut, then the second and third clauses are equivalent to
the classical formulas
\begin{tabbing}
\Indent{} \= \Indent{} \= \kill
\>  $p(b, y) \leftarrow q(y) \And r(y).$ \\
\>  $p(x, y) \leftarrow (\Not(x=b) \Or (x=b \And \Not q(y))).$
\end{tabbing}
However, we lose the ability to select the first solution with soft cut.

A construct related to cut is the ``if-then-else'' construct,
usually written $(G_1$ \verb/->/ $G_2 ; G_3)$ and read ``if $G_1$ then $G_2$
else $G_3$''.  This construct is often syntactic sugar for a
hard-cut-like operation; that is, the evaluation of
$(G_1$ \verb/->/ $G_2 ; G_3)$ is equivalent to the evaluation of a goal
$p(x_1, \ldots, x_n)$ against the program
\begin{tabbing}
\Indent{} \= \kill
\>  $p(x_1, \ldots, x_n) \colondash G_1, !, G_2$ \\
\>  $p(x_1, \ldots, x_n) \colondash G_3$ \\
\end{tabbing} \vspace{-1ex}
where $x_1, \ldots, x_n$ are the free variables in $G_1, G_2, G_3$.

The cut in the if-then-else construct is hard cut in most Prologs.
The choice construct of the Mercury language \cite{mercury-jlp}
is written in this way and uses soft cut; Mercury has no other
choice construct.

\subsection{Semantics of Prolog and Cut}

The least-model semantics \cite{vanE-kow} is traditionally
viewed as the standard one for pure logic programming as it was
originally conceived.  However, the depth-first search of Prolog
and similar systems makes it difficult to fit them into the
least-model framework, at least if we want a semantics with
respect to which the system is sound and complete.  Evidently
some other form of semantics is needed to characterize
depth-first logic programming systems precisely, whether taking
cut into consideration or not.

The {\it operational} semantics of Prolog with cut was not
formally defined in a self-contained system
until Billaud's 1990 paper \cite{billaud-cut-tcs}.  In
Billaud's semantics, when a predicate is called, the current
backtrack stack is stored; the
execution of a cut corresponds to discarding the current
backtrack stack and replacing it with
the one stored by the current predicate.

Various authors have given {\it denotational} semantics for
Prolog with cut
\cite{prolog-continuation,boerger-opsem,baudinet-jlp},
including Billaud in his original paper \cite{billaud-cut-tcs}.
Some of these approaches have proven equivalence with an
operational semantics.
These papers were based on earlier work in operational and
denotational semantics of Prolog, including
\cite{jones-mycroft,deransart-std-inria,denotationalization,%
debray-denotational,nicholson-foo}.

The denotational approaches essentially view a Prolog program as
a function from goals to sequences of answer substitutions, and
``reify'' notions like unification and answer substitution
sequence by giving abstract mathematical constructs
corresponding to them.  Such approaches are able to handle any
operational model which transforms a goal into a sequence of
substitutions using unification.  This includes models with any
conceivable sound or unsound strategy for negation and cut; for
instance, sound soft cut, unsound negation as failure, or a
negation operator which judges $\Not p(t)$ to be true iff $t$
unifies with 42.  Therefore, although a reifying semantics may
be very useful for some purposes (for instance, to use as a
guide for implementation of a standard computational model), the
existence of such a semantics does not by itself suggest that
the system thus characterized is any more than an operational
superset (or superset of a subset) of pure logic programming.

In contrast, what may be called the ``non-reifying'' semantic
tradition
\cite{andrews-phd-dd,andrews-lnaf-tcs,staerk-lptp-jlp,elbl-jlp-1999}
gives characterizations of the success and failure of Prolog
goals not involving reified answer substitutions and
unification.  Andrews' earliest characterizations
\cite{andrews-phd-dd} took account only of depth-first Prolog
without builtins, negation or cut.  Andrews
\cite{andrews-lnaf-tcs} and \Staerk{} \cite{staerk-lptp-jlp}
then extended this to systems with negation as failure, Andrews
by characterizing floundering and \Staerk{} by imposing a mode
restriction.  More recently, Elbl \cite{elbl-jlp-1999} has given
a semantics for depth-first logic programming which uses more
abstract denotations to achieve compositionality, and extends
this semantics to take account of negation with a similar mode
restriction to \Staerk{}'s.

These more logical approaches draw their power from
expressing the semantics of Prolog in a manner which allows
them to avoid encoding operational notions such as unification
into the semantics.  Without such a property, proofs using
\Staerk's proof assistant \cite{staerk-lptp-jlp}, for instance,
would have to reason about unification at almost every step.

We should note that even reifying semantics can act as the basis
of powerful theorem provers if they are automated.  For example,
Lindenstrauss, Sagiv and Serebrenik
\cite{lindenstrauss-auto-term-iclp97,lindenstrauss-termilog-cav97}
discuss automatic proofs of strong termination based on
term rewriting techniques.  However, in proving termination and
(especially) correctness properties, it is often necessary to
have human intervention, in order to deduce generalizations to
be proven by induction or norms for proving termination.

\subsection{Termination}

We seek an abstract semantics with respect to which some large
subset of Prolog with cut is sound {\it and complete}.  The
soundness property allows us to argue that any outcome which a
Prolog goal does return is consistent with the semantics.  The
completeness property, however, allows us to argue that the
semantics does not judge a goal to be true (resp.\ false)
unless it actually succeeds (resp.\ fails) according to the
operational semantics; that is, that we have
precisely captured {\it termination} of goals.  We must
therefore define exactly what we mean by termination of a goal.
In this paper, we study {\it left-to-right} termination, which
subsumes the more widely-studied notion of {\it strong} termination.

A Prolog query can have one of several outcomes.
It can succeed or fail, or it can diverge (fail to
terminate altogether).  If a query succeeds, Prolog typically
gives us the option of finding more solutions.  If we keep asking
for more solutions, there are three things that may happen:
the query may eventually
fail back to the top level and report
no more solutions; the query may return a finite number of
solutions and then diverge; or the query may return
an infinite number of solutions.
We may label these outcomes as:
\begin{enumerate}
\item Success:
  \begin{enumerate}
  \item Finite number of solutions, then failure.
  \item Finite number of solutions, then divergence.
  \item Infinite number of solutions.
  \end{enumerate}
\item Failure.
\item Divergence.
\end{enumerate}

These outcomes correspond to the shape of the resolution
search tree for systems with a left-to-right subgoal selection
rule, and the placement of solutions within that tree.  (In the
following, we assume that the leftmost subgoal is always selected,
that the children of each node of the search tree
correspond, left to right, to the sequence of clauses defining
the selected subgoal's predicate, and that the search rule is
also left-to-right.)  If the
tree is finite, we get outcome 1(a) or 2.  If it has some infinite
path, and there is a finite number of solutions to the left
of the leftmost infinite path, we get outcome 1(b) or 3. 
Otherwise, there is an infinite number of solutions to the left
of the leftmost infinite path (outcome 1(c)), and we can obtain
only a finite prefix of the sequence of solutions by
backtracking.

The two kinds of termination most often mentioned in the
literature are {\it existential termination} and {\it universal
termination}.  A query existentially terminates either if it
fails, or if there is a solution somewhere in the search tree.
Knowing that a query existentially terminates is thus useful
primarily if we are studying breadth-first implementations or
nondeterministic operational semantics.
A query universally terminates if the search tree is finite (i.e.,
a search on any path terminates).  Universal termination
therefore corresponds only to cases 1(a) and 2 above.

Most of the work on proving termination of Prolog programs (e.g.,
\cite{pluemer-lncs,apt-strong-info-comp,bezem-strong-jlp,apt-marchiori,staerk-lptp-jlp}%
) has
concentrated on universal termination.  Because of our interest
in features of practical logic programming systems such as
Prolog, in this paper we continue to study
what we refer to as {\it depth-first termination}.  A query
depth-first terminates if it returns at least one solution, or
if it fails.  Depth-first termination thus encompasses outcomes
1(a)-(c) and 2 above, and thus identifies a larger set of
queries as terminating than universal termination.  It also
corresponds to one of a Prolog user's intuitive notions of termination
of a goal.

Depth-first termination is what we will have to characterize if
we want to take account of the behaviour of cut.  Cut cuts
away all but the first solution returned from the portion of the
clause before the cut, so all that is important to the semantics
is that the portion before the cut returns at least one solution
or fails.  Note, however, that even in the absence of cut, a
goal formula $G$ universally terminates iff the query
$(G \And false)$ (in Prolog parlance, {\tt (G, fail)})
depth-first terminates.  Depth-first termination is thus
strictly more general than universal termination.

\section{Notation and Syntax of Extended Programs}

\newcommand{\Bar}{~~|~~}

In this section, we define the syntax of programs that we will
use for the rest of the paper.  It is a generalization of the
subset of Prolog including cut (!), negation as failure, and
defined predicates.  It does not include problematic built-in
predicates such as \verb/assert/ and \verb/retract/, \verb/var/,
\verb/nonvar/, and \verb/setof/, each of which merits further
study but whose inclusion might confuse the issues we study here.

We use the following meta-variables:
$B$, $C$, $F$, $G$ and $H$ for formulas,
$s$ and $t$ for terms, and
$x$, $y$ and $z$ for variables,
all possibly primed or subscripted.
We use $\vec{x}$, $\vec{t}$, etc.\ generally to stand for
sequences of variables, terms, etc.
We use $\exists \vec{x}$ as notation to stand for
$\exists x_1 \ldots \exists x_n$, where
$\vec{x} = (x_1, \ldots, x_n)$.

We define an extended notion of {\it goal formula} (or simply
{\it formula}), representing a query or an element of a clause body.
The BNF definition of a formula is as follows.\vspace{2mm} \\
\begin{tabular}{rrl}
$G$ & ::= &
      $(t = t)  \Bar  p(t, \ldots, t)  \Bar  G \And G  \Bar  G \Or G$ \\
  & $|$ & $\Not G  \Bar  \exists x~G  \Bar  if[\vec{x}](G, G)$ \\
\end{tabular} \\
All the connectives are standard except the $if$ connective.
$if[\vec{x}](B,C)$ is a variable binding construct, which binds
all the variables in the list $\vec{x}$.
$if[\vec{x}](B, C)$ is computed as follows: if
$\exists \vec{x}(B)$ is false, so is
$if[\vec{x}](B, C)$; otherwise,
$if[\vec{x}](B, C)$ is equivalent to
$C\theta$, where $\theta$ is the first substitution
for $\vec{x}$ returned by the computation of $B$.
This form of formula allows us to express a Prolog
program with cuts in a ``completed'' form (see section
\ref{completions-section}).

\label{true-false-section}
We assume a standard syntax of terms.  We assume that the
language of the program contains at least two terms, which we
will refer to as $0$ and $1$.  We define the formula $true$ as
$0=0$, and the formula $false$ as $0=1$.

Because we will be speaking of clauses with cut, we cannot 
use the standard logic-programming definition of clause.
The BNF definitions of formula, clause, clause body, and clause
body element used in this paper are as follows.\vspace{2mm} \\
\begin{tabular}{lll}
$clause$ & ::= & $p(t, \ldots, t) \colondash body$ \\
$body$ & ::= &
      $\epsilon  \Bar  bodyelt, body$ \\
$bodyelt$ & ::= &
      $G  \Bar  !$ \\
\end{tabular} \\
($\epsilon$ is the empty expression.)
As in Prolog, we generally write a clause of the form
$p(t_1, \ldots, t_n) \colondash \epsilon$
as simply $p(t_1, \ldots, t_n)$.
Note that we restrict the cut to occurring ``at the top level''
in clauses.  In most Prologs it is possible to use cut within a
complex formula (for instance, a disjunction), but such cuts are
seldom used and their effect is generally said to be
undefined\footnote{
Billaud's operational semantics of cut \cite{billaud-cut-tcs}
defines a behaviour of cuts within complex formulas which is
consistent with the operational semantics of some Prolog interpreters.
}.

A {\it program} is a sequence of clauses.
It is clear that the syntax of programs, as defined here,
generalizes the syntax of Prolog programs with only
literals and cuts as body elements.
For simplicity, we assume that each predicate is defined
with a distinct arity in a given program; that is, that at every
occurrence of a predicate name, it is given the same number of
parameters.  We say that a clause {\it defines} predicate $p$ if
the head of the clause has predicate $p$.
We use $clauses(p,P)$ to stand for the sequence of
clauses defining predicate $p$ in program $P$.

\label{example-section}
As an example of a program in the extended syntax, consider the
following standard definition of a ``delete'' predicate:
\begin{tabbing}
\Indent{} \= \kill
\>  $d(x, [~], [~])$ \\
\>  $d(x, [x|ys], zs) \colondash !, d(x, ys, zs)$ \\
\>  $d(x, [y|ys], [y|zs]) \colondash d(x, ys, zs)$
\end{tabbing}
The goal $d(x, y, z)$ deletes all occurences of the element $x$
in the list $y$, resulting in the list $z$.
As we will see, the following definition is equivalent:
\begin{tabbing}
\Indent{} \= $\Or$ \= \kill
\>  $d(x, y, z) \colondash$ \\
\>  \>  $(y=[~] ~ \And ~ z=[~])$ \\
\>  $\Or$ \> $if[ys](y=[x|ys], d(x, ys, z))$ \\
\>  $\Or$ \> $(\Not\exists ys(y=[x|ys]) ~ \And$ \\
\>        \> \Indent{} $\exists y' \exists ys \exists zs(y=[y'|ys] \And z=[y'|zs] \And d(x, ys, zs)))$
\end{tabbing}

\section{The Liberal Operational Semantics}
\label{opsem-section}

In order to define precisely the logic programming systems
which will be the focus of our study, we must define precisely
their operational, or procedural, semantics.
In this section, we define two operational semantics
(the second simpler than the first) for
the extended logic programs defined in the last section.
Because they share Prolog's rather lax, non-logical interpretation
of negation and cut, they are referred to as ``liberal''
operational semantics.  The second of these semantics will be
used as the basis of the more ``conservative'' semantics of
the next section, which regains the witness properties.

Traditionally,
operational semantics of logic programming are given using
variants of resolution, in particular SLD-resolution.
However, in the presence of such features as depth-first search,
negation as failure and cut, SLD-resolution-based operational semantics
require an additional superstructure of definitions, for instance
to define the order in which branches of the SLD-tree are searched.
We therefore follow other researchers
\cite{deransart-std-inria,billaud-cut-tcs} in defining operational
semantics for our system using the style which has come to be
known as SOS, or Structured Operational Semantics \cite{plotkin-opsem}.

The rules in this paper are presented in groups,
which (following \cite{abadi-cardelli-objects}) are referred to
as ``fragments'', to emphasize that they are only parts
of formal systems.  We define various different
operational semantics for various different purposes; each semantics will
be made up of several of these fragments.

In this section, we first present some basic definitions in section
\ref{defns-subsection}.  In section \ref{libgen-subsection},
we define the ``liberal general'' operational semantics.
This semantics takes its name from its liberal attitude and
the fact that it can handle general programs
(with multi-clause definitions and cut).

Traditional Prolog multi-clause predicate definitions
turn out to be awkward to work with in the presence of cut.
Predicates defined with a single clause are more convenient to
work with; but is it always possible to transform a program
with multi-clause definitions into one with single-clause
definitions?  In section \ref{completions-section} we answer this
question in the affirmative, defining a ``completed form''
for programs and giving an algorithm which transforms a program
to completed form.
In section
\ref{libcomp-subsection}, we give the ``liberal completed''
semantics, which is defined only for completed-form programs and is much
simpler than the liberal general semantics.  It is this
liberal completed semantics that we use as the basis of the
safer, ``conservative'' semantics of the rest of the paper.

Finally, in section \ref{inadequate-section}, we show formally
that the liberal semantics, like the Prolog systems they
characterize, are problematic from a logic programming point of
view because they violate not only logic, but also the
weaker witness properties.

\subsection{Basic Definitions}
\label{defns-subsection}

This section defines some basic notions of the operational
semantics, namely goal stacks, results, judgments and
computations.

The judgements of the operational semantics contain
{\it goal stack elements} and {\it results}.
A goal stack element represents a subgoal to solve,
possibly with information about how to solve it.  A goal stack
element can be one of the following:
\begin{itemize}
\item a formula;
\item an expression of the form $p(t_1, \ldots, t_n) using(\gamma)$,
  where $\gamma$ is a sequence of clauses; or
\item an expression of the form $body(\eta)$, where $\eta$ is a
  clause body (i.e., a possibly empty sequence of body elements).
\end{itemize}
A goal stack element of the form $p(t_1, \ldots, t_n) using(\gamma)$
represents a predicate call along with the sequence of
clauses remaining to be used in its processing; a goal stack element
of the form $body(\eta)$
represents a predicate body, possibly containing cuts.  (We
distinguish a predicate body from a regular sequence of formulas
in this way because a body with cuts demands some special
treatment.)  We define a {\it goal stack} as a sequence of goal
stack elements.

In this paper, the {\it result} of a computation in the
operational semantics can be one of four things:
\begin{itemize}
\item A substitution $\theta$, indicating a successful computation
  returning $\theta$ as the solution;
\item $fail$, indicating failure to find a substitution;
\item $flounder$, indicating that a mode restriction
  has been violated (see Section \ref{conservative-section}); or
\item $diverge$, indicating that the operational semantics
  believes the computation to diverge (see Section \ref{sems-section}).
\end{itemize}
Only the first two results are possible with the semantics in
this section, but the others will be possible in later
semantics.

A {\it judgement} of an operational semantics is an expression
of the form $(\theta: \alpha \Goes{P} \rho)$, where $\theta$ is a
(finite representation of a)
substitution, $\alpha$ is a goal stack containing no free
variables in the domain of $\theta$,
$P$ is a program, and $\rho$ is a result.  A judgement indicates
that the computation of the goals in $\alpha$, under the current
substitution $\theta$ and the program $P$, has the result $\rho$.

A {\it computation} in a given operational semantics
is a tree, written root-down, in which each
node is a judgement, and where the relationship between each
node and its children is defined by the rules in that
operational semantics.  Computing the outcome of a Prolog
goal $G$ with respect to program $P$ corresponds to finding a
result $\rho$ and a
computation whose root node is $((): G \Goes{P} \rho)$, where
$()$ is the empty substitution.  Generally, we will drop the
$P$ subscript where its value is clear.

In the operational semantics, we use $\alpha$ to stand for a
goal stack, and $\eta$ to stand for a sequence of
body elements.  We use $\gamma$ to stand for a sequence of
clauses; to distinguish sequences of clauses more clearly from
sequences of goal stack or body elements, we separate clauses
in a sequence by semicolons, and goal stack or body elements
by commas.

\subsection{The Liberal General Semantics}
\label{libgen-subsection}

\begin{figure}[tp]

\begin{center} 
\begin{tabular}{ccc}
~
  & ~~~ &
  $\overline{\theta'': \epsilon \Goes{} \theta''}$ \\
~
  & ~~~ &
  $\overline{\theta'[x':=a]: z=[~] \Goes{} \theta''}$ \\
~
  & ~~~ &
  $\overline{\theta'[x':=a]: [~]=[~], z=[~] \Goes{} \theta''}$ \\
$\overline{\theta': \epsilon\Goes{} \theta'}$
  & ~~~ &
  $\overline{\theta': a=x', [~]=[~], z=[~] \Goes{} \theta''}$ \\
$\overline{[x:=a,ys:=[~]]: z=zs \Goes{} \theta'}$
  & ~~~ &
  $\overline{\underline{\theta': d(a,[~],z)using(C_1;C_2;C_3) \Goes{} \theta''}}$ \\
$\overline{[x:=a]: [a]=[a|ys], z=zs \Goes{} \theta'}$
  & ~~~ &
  $\theta': d(a,[~],z) \Goes{} \theta''$ \\
$\overline{(): a=x, [a]=[x|ys], z=zs \Goes{} \theta'}$
  & ~~~ &
  $\overline{\theta': body(d(a,[~],z)) \Goes{} \theta''}$ \\
\cline{1-3}
\multicolumn{3}{c}{
    $(): d(a,[a], z)using(C_2;C_3) \Goes{} \theta''$
  } \\
\end{tabular}
\end{center}

\begin{center} 
\begin{tabular}{ccc}
$\overline{[x:=a]: [a]=[~], z=[~] \Goes{} fail}$
  & ~~~ &
  (see above) \\
$\overline{(): a=x, [a]=[~], z=[~] \Goes{} fail}$
  & ~~~ &
  $\overline{(): d(a,[a], z)using(C_2;C_3) \Goes{} \theta''}$ \\
\cline{1-3}
\multicolumn{3}{c}{
  $\underline{(): d(a,[a], z)using(C_1;C_2;C_3) \Goes{} \theta''}$
} \\
\multicolumn{3}{c}{
  $(): d(a,[a], z) \Goes{} \theta''$
} \\
\end{tabular}
\end{center}

\caption{An example computation in the liberal general semantics
  with respect to the first ``delete'' program of Section 3.
  The computation is split into two pieces in order to fit on the page.
}
\label{libgen-example-fig}
\end{figure}

\begin{figure}[tp]

\noindent
\Rule{Unif/succ:}
  {\theta\sigma: \alpha\sigma \Goes{} \rho}
  {\theta: (s = t), \alpha \Goes{} \rho}
  {where $\sigma$ is an mgu of $s$ and $t$}
\\ %
\Rule{Unif/fail:}
  {}
  {\theta: (s = t), \alpha \Goes{} fail}
  {where $s$ and $t$ are not unifiable}
\\ %
\Rule{Success:}
  {}
  {\theta: \epsilon \Goes{} \theta}
  {}
\\ %
\Rule{Conj:}
  {\theta: B, C, \alpha \Goes{} \rho}
  {\theta: B \And C, \alpha \Goes{} \rho}
  {}
\\ %
\Rule{Disj/nofail:}
  {\theta: B, \alpha \Goes{} \rho}
  {\theta: B \Or C, \alpha \Goes{} \rho}
  {where $\rho$ is not $fail$}
\\ %
\Rule{Disj/fail:}
  {\theta: B, \alpha \Goes{} fail  \Sep
   \theta: C, \alpha \Goes{} \rho}
  {\theta: B \Or C, \alpha \Goes{} \rho}
  {}
\\ %
\Rule{Exists:}
  {\theta: B[x:=x'], \alpha \Goes{} \rho}
  {\theta: \exists x(B), \alpha \Goes{} \rho}
  {where $x'$ does not occur in the conclusion}

\caption{
  [Basic], the operational semantics rules fragment for the
  basic logic programming connectives.
}
\label{basic-rules-fig}
\end{figure}

\begin{figure}[tp]

\noindent
\Rule{Not/succ:}
  {\theta: B \Goes{} \theta'}
  {\theta: \Not B, \alpha \Goes{} fail}
  {}
\\ %
\Rule{Not/fail:}
  {\theta: B \Goes{} fail  \Sep  \theta: \alpha \Goes{} \rho}
  {\theta: \Not B, \alpha \Goes{} \rho}
  {}
\\ %
\Rule{If/succ:}
  {\theta: B[\vec{x}:=\vec{x}'] \Goes{} \theta'  \Sep
   \theta': C[\vec{x}:=\vec{x}']\theta', \alpha\theta' \Goes{} \rho}
  {\theta: if[\vec{x}](B,C), \alpha \Goes{} \rho}
  {where $\vec{x}'$ do not appear in the conclusion}
\\ %
\Rule{If/fail:}
  {\theta: B[\vec{x}:=\vec{x}'] \Goes{} fail}
  {\theta: if[\vec{x}](B,C), \alpha \Goes{} fail}
  {where $\vec{x}'$ do not appear in the conclusion}

\caption{
  [Liberal Choice], the operational semantics rules fragment for
  dealing with ``not'' and ``if'' in a liberal manner.
}
\label{lib-choice-rules-fig}
\end{figure}

\begin{figure}[tp]

\noindent
\Rule{Pred:}
  {\theta: p(t_1, \ldots, t_n) using (\gamma), \alpha \Goes{} \rho}
  {\theta: p(t_1, \ldots, t_n), \alpha \Goes{} \rho}
  {where $\gamma$ is $clauses(p,P)$, renamed apart from any
   free variables in the conclusion}
\\ %
\Rule{Using/cut/succ:}
  {\theta: s_1=t_1, \ldots, s_n=t_n, \eta_1 \Goes{} \theta'  \Sep
   \theta': body(\eta_2)\theta', \alpha\theta' \Goes{} \rho}
  {\theta: p(s_1, \ldots, s_n) using (C,\gamma), \alpha \Goes{} \rho}
  {where $C$ is of the form $p(t_1, \ldots, t_n) \colondash \eta_1, !, \eta_2$,
   and $\eta_1$ contains no cuts}
\\ %
\Rule{Using/cut/fail:}
  {\theta: s_1=t_1, \ldots, s_n=t_n, \eta_1 \Goes{} fail  \Sep
   \theta: p(s_1, \ldots, s_n) using (\gamma), \alpha \Goes{} \rho}
  {\theta: p(s_1, \ldots, s_n) using (C,\gamma), \alpha \Goes{} \rho}
  {where $C$ is of the form $p(t_1, \ldots, t_n) \colondash \eta_1, !, \eta_2$,
   and $\eta_1$ contains no cuts}
\\ %
\Rule{Using/nocut/succ:}
  {\theta: s_1=t_1, \ldots, s_n=t_n, \eta, \alpha \Goes{} \theta'}
  {\theta: p(s_1, \ldots, s_n) using (C,\gamma), \alpha \Goes{} \theta'}
  {where $C$ is of the form $p(t_1, \ldots, t_n) \colondash \eta$,
   and $\eta$ contains no cuts}
\\ %
\Rule{Using/nocut/fail:}
  {\theta: s_1=t_1, \ldots, s_n=t_n, \eta, \alpha \Goes{} fail  \Sep
   \theta: p(s_1, \ldots, s_n) using (\gamma), \alpha \Goes{} \rho}
  {\theta: p(s_1, \ldots, s_n) using (C,\gamma), \alpha \Goes{} \rho}
  {where $C$ is of the form $p(t_1, \ldots, t_n) \colondash \eta$,
   and $\eta$ contains no cuts}
\\ %
\Rule{Using/empty:}
  {}
  {\theta: p(s_1, \ldots, s_n) using (\epsilon), \alpha \Goes{} fail}
  {}

%
%
%

\Rule{Body/cut/succ:}
 {\theta: \eta_1 \Goes{} \theta'  \Sep
  \theta': body(\eta_2)\theta', \alpha\theta' \Goes{} \rho}
 {\theta: body(\eta_1, !, \eta_2), \alpha \Goes{} \rho}
 {where $\eta_1$ contains no cuts}
\\ %
\Rule{Body/cut/fail:}
 {\theta: \eta_1 \Goes{} fail}
 {\theta: body(\eta_1, !, \eta_2), \alpha \Goes{} fail}
 {where $\eta_1$ contains no cuts}
\\ %
\Rule{Body/nocut:}
 {\theta: \eta, \alpha \Goes{} \rho}
 {\theta: body(\eta), \alpha \Goes{} \rho}
 {where $\eta$ contains no cuts}

\caption{
  [General Predicates], the operational semantics
  rules fragment for dealing with general (multi-clause) predicate
  definitions.
}
\label{general-pred-rules-figb}
\end{figure}

The first operational semantics we study, as described
above, is the {\it liberal general} semantics.
It is made up of the fragments
[Basic] (Figure \ref{basic-rules-fig}),
[Liberal Choice] (Figure \ref{lib-choice-rules-fig}), and
[General Predicates] (Figure
\ref{general-pred-rules-figb}).
The liberal general semantics corresponds
to most common implementations of Prolog, which employ hard cut and
unsound negation as failure.
We begin this section by looking at an example computation,
and then discuss the individual rules of the liberal general
semantics in more detail.

\subsubsection{Example Computation}
\label{example-comp-subsection}

Figure \ref{libgen-example-fig} shows an example computation in
the liberal general semantics.
(The clauses $C_1, C_2, C_3$ are the clauses for $d$
from the three-clause version defined in Section
\ref{example-section}).
The substitution $\theta'$ is $[x:=a,ys:=[~],zs:=z]$,
and
the substitution $\theta''$ is $[x:=a,ys:=[~],zs:=[~],x':=a,z=[~]]$.)
This computation, like all
computations, gives the result of the computation within the
same judgement as the original goal.  Therefore it may not be
clear how to obtain a result from knowing only the goal we want
to solve.  The example illustrates how we can do so
in a systematic fashion by applying rules bottom-up.

We start (at the bottom) with
the goal formula $d(a, [a], z)$ and the empty substitution; our
task is to determine the result expression, to the right of
the $\Goes{}$ symbol.  Since $d(a,[a],z)$ is a predicate
call, we know that the bottommost rule is a Pred rule,
that the substitution in the premise is still empty, and that
the goal stack in the premise is
$d(a,[a], z)using(C_1;C_2;C_3)$.
We therefore apply that rule at the bottom of the computation.
We have now reduced the problem of finding the result of
$((): d(a, [a], z))$ to that of finding the result of
$((): d(a,[a], z)using(C_1;C_2;C_3))$.

At this point, we can apply either the Using/nocut/succ or the
Using/nocut/fail rule; we do not know which is applicable.
However, we know that if Using/nocut/succ is applicable,
the substitution in the left-hand premise is the
empty substitution and the goal stack in the left-hand premise is
$(a=x, [a]=[~], z=[~])$; we also know that if Using/nocut/fail
is applicable, then the substitution in the (only) premise
is again empty and the goal stack in the premise is again
$(a=x, [a]=[~], z=[~])$.  If the result of this goal stack is
$fail$, then Using/nocut/fail is applicable; if it is some
substitution $\theta$, then Using/nocut/succ is applicable.
We therefore choose as our next task to find the result of
$((): a=x, [a]=[~], z=[~])$.

As it turns out, in two simple steps (a Unif/succ step and
a Unif/fail step) we can determine that
$((): a=x, [a]=[~], z=[~] \Goes{} fail)$.  Therefore
we choose Using/nocut/fail as the rule to apply.  This
choice determines the form of the substitution (again, the
empty substitution) and the goal stack
($d(a,[a], z)using(C_2;C_3)$) in the right-hand premise.
We can repeat this process of finding results
in order to obtain the result $\theta''$ of
$d(a,[a], z)using(C_2;C_3)$, which is inherited by our
original goal $d(a, [a], z)$ as its result.
$\theta''$ contains the mapping $[z:=[~]]$; thus the
computation has correctly told us that the result of
deleting $a$ from the list $[a]$ is the empty list.

In general, whenever we are faced with a choice of two rules,
the above strategy will work.  The form of substitution and goal
stack in one of the premises can be uniquely determined, and the
choice of rule and form of substitution and goal stack in the
other premise (if another is needed) can be uniquely determined
from the result of the
first premise.  Thus, information in a computation
can be seen as ``flowing'' in a clockwise manner around the
perimeter of the computation.

\subsubsection{The Rules}
\label{rules-subsection}

We now describe the general significance of the rules in the
liberal general semantics in terms of how the different kinds of goal
stack elements are handled.

The equality rules in the [Basic] fragment describe the usual
results of unification; if unification fails, the entire goal
stack fails, but if it succeeds, the computation proceeds under
the mgu.
The first order connective rules in [Basic] express the usual
operation of Prolog interpreters.  We solve a conjunction by
solving each of its conjuncts in turn, left to right.  We solve
a disjunction by attempting to solve its left-hand disjunct and
the rest of the subgoals; if this is solvable, we can ignore the
right-hand disjunct, but if not, we attempt to solve that
disjunct with the rest of the subgoals.  Finally, we solve an
existential formula (corresponding to a free variable in a
clause) by renaming its variable apart from the rest of the
variables in the goal.

In the [Liberal Choice] rules, we solve a negation by solving
the negated formula, inverting the sense of the result at the end.
This is the usual unsound strategy, which will be corrected
in the system with firm cut.
Similarly, the formula
$if[\vec{x}](B, C)$ is computed by first computing $B$ and
checking the result.  If the result is a successful computation
returning satisfying substitution $\theta$, then $\theta$ is
used to compute $C$; otherwise, the whole formula fails.
This will also be modified in the system with firm cut,
in order to achieve the witness properties.

The predicate call and clause selection rules of the
[General Predicates] fragment reflect how Prolog
backtracks over clauses and cuts away alternate solutions.
We ``launch'' the processing of a predicate
call by collecting the clauses in the program defining the predicate
into an initial $using$ expression.
Then, if the first clause contains a cut, we process first only
the part before
the cut.  On success, we retain the substitution returned and
discard the other clauses, but on failure,
we discard that first clause and repeat the procedure.
This characterizes the behaviour of
Prolog clauses with cut.

Conversely, if the first clause does not contain
a cut, we process the entire clause body {\it along with the rest
of the subgoals}.  Again, on success of the goal stack we
discard the other clauses, and on failure of the goal stack, the
first clause.
However, because we have included the rest of the subgoals in the
goal stack,
we retain the option of returning to another clause if a
subgoal fails later in the computation.  This characterizes the
behaviour of usual Prolog clauses without cut.

Finally, the predicate body rules reflect how cuts in a clause body
after the first cut may prune the search tree.  If a clause body
has cuts, then the portion before the first cut
is processed first; if it returns a solution, we process the rest
of the body with that first solution, and otherwise the entire body fails.
If the body has no cuts, however, it is processed just as a
sequence of formulas.

\subsection{Completed Forms of Programs}
\label{completions-section}

In this section, we show that it is possible to transform any
program into one in a ``completed'' form, in which every
predicate is defined by a single clause without cuts.  This is
valuable because programs in completed form are much easier to
work with in the proofs we need to do.  We begin by giving the
transformation algorithm, show an example of how it transforms a
program, and then prove the required properties of the
transformation algorithm.

We say that a program is {\it in completed form} when each of
the following conditions hold:
\begin{enumerate}
\item The parameters in the clause head are distinct variables;
\item There is only one clause defining each predicate;
\item The body of each clause consists of a single formula; and
\item The free variables in the body are a subset of the parameters
  in the head.
\end{enumerate}
Our transformation of programs into completed forms depends on
the fact that our definition of formula includes the $if$
connective, which allows us to achieve the effect of cuts;
in fact, this is the main reason why $if$ was included in the
syntax and operational semantics of our language.

\subsubsection{Transformation Algorithm}
\label{transformation-subsection}

Here, we give an algorithm which progressively
transforms a program into completed form, by replacing clauses
with other clauses.  The program, as it is being transformed,
will progressively satisfy each of the following properties.
\begin{itemize}
\item[(A)] The parameters in the clause head are distinct variables.
\item[(B)] Each clause body begins and ends with a formula, and
  alternates formulas and cuts.
\item[(C)] Each clause has at most one cut; that is, each clause
  body consists of either a singleton formula $F$, or a sequence
  $F, !, G$.
\item[(D)] The last clause defining each predicate has a body
  which is a single formula, having no free variables except
  those appearing in the head.
\item[(E)] Each predicate is defined by exactly one clause.
\end{itemize}

The algorithm is as follows.
\begin{itemize}
\item[1.] Choose a countable
  sequence of variables not appearing in the program.
  We will refer to these variables as $x_1, x_2, \ldots$
  in the rest of the algorithm.
\item[2.] While there is some clause in the program not of
  the form $(p(x_1, \ldots, x_n) \colondash \eta)$:
  \begin{itemize}
  \item[2.1.] Choose one such clause $C$, of the form \\
    $p(t_1, \ldots, t_{k-1}, t_k, x_{k+1}, \ldots, x_n) \colondash \eta$,
    where $t_k$ is not $x_k$.
  \item[2.2.] If $t_k$ is a variable $y$ distinct from
    $x_1, \ldots, x_n$, then replace $C$ in the program by $C[y:=x_k]$.
  \item[2.3.] Otherwise, replace $C$ by \\
    $p(t_1, \ldots, t_{k-1}, x_k, x_{k+1}, \ldots, x_n) \colondash
    (x_k = t_k), \eta$.
  \end{itemize}
  (After this while loop has been completed,
  we can assume that property (A) above is satisfied.)
\item[3.] While there is some clause of the form
  $p(x_1, \ldots, x_n) \colondash \eta_1, F, G, \eta_2$, where $F$
  and $G$ are formulas:
  choose one such clause and transform it to the
    form $p(x_1, \ldots, x_n) \colondash \eta_1, (F \And G), \eta_2$.
\item[4.] While there is some clause with an empty body:
  choose one such clause and replace the body by the single
  formula $true$ (i.e., $0=0$).
\item[5.] While there is some clause with two consecutive cuts:
  choose one such clause and replace the consecutive cuts by a
  single cut.
\item[6.] While there is some clause beginning with a cut:
  choose one such clause and insert the formula $true$ before
  the first cut.
\item[7.] While there is some clause ending with a cut:
  choose one such clause and insert the formula $true$ after
  the last cut.
  (We can now assume that property (B) above is satisfied.)
\item[8.] While there is some clause of the form
  $p(x_1, \ldots, x_n) \colondash \eta, !, F, !, G$:
  \begin{itemize}
  \item[8.1.] Select one such clause.
  \item[8.2.] Select a predicate name $q$ not appearing in the
    program.
  \item[8.3.] Add a clause to the program of the form
    $q(\vec{y}) \colondash F, !, G$, where
    $\vec{y}$ are all the free variables of $F, G$.
  \item[8.4.] Replace the original selected clause by 
    $p(x_1, \ldots, x_n) \colondash \eta, !, q(\vec{y})$.
  \end{itemize}
  (We can now assume that property (C) above is satisfied.)
\item[9.]
  Repeat until the last clause of all predicates is
  of the form $p(x_1, \ldots, x_n) \colondash G$,
  where all free variables of $G$ appear in the head:
  \begin{itemize}
  \item[9.1.] Choose the last clause of one predicate for which
    this is not the case; let it be of the form
    $p(x_1, \ldots, x_n) \colondash \eta$.
  \item[9.2.] If $\eta$ is some singleton formula $G$, replace the
    clause by \linebreak[4]
    $p(x_1, \ldots, x_n) \colondash \exists \vec{y} (G)$,
    where $\vec{y}$ are all the free variables of $G$ not
    in $x_1, \ldots, x_n$.
  \item[9.3.] Otherwise, $\eta$ is a sequence of the form $F, !, G$.
    Replace the clause by
    $p(x_1, \ldots, x_n) \colondash if[\vec{y}](F, G)$,
    where $\vec{y}$ are all the free variables of $F, G$ not
    in $x_1, \ldots, x_n$.
  \end{itemize}
  (We can now assume that property (D) above is satisfied.)
\item[10.] While there is some predicate which is defined by more
  than one clause:
  \begin{itemize}
  \item[10.1.] Choose one such predicate $p$.  Let the second-last
    clause defining $p$ be
    $p(x_1, \ldots, x_n) \colondash \eta$, and let the last clause
    defining $p$ be
    $p(x_1, \ldots, x_n) \colondash H$.
  \item[10.2.] If $\eta$ is some singleton formula $G$, replace the
    two clauses by the single clause
    $p(x_1, \ldots, x_n) \colondash \exists \vec{y} (G) \Or H$,
    where $\vec{y}$ are all the free variables of $G$ not
    in $x_1, \ldots, x_n$.
  \item[10.3.] Otherwise, $\eta$ is a sequence of the form $F, !, G$.
    Replace the two clauses by the single clause
    $p(x_1, \ldots, x_n) \colondash
       if[\vec{y}](F, G) \Or ((\Not\exists \vec{y} (F)) \And H)$,
    where $\vec{y}$ are all the free variables of $F,G$ not
    in $x_1, \ldots, x_n$.
  \end{itemize}
  (We can now assume that property (E) above is satisfied.)
\end{itemize}

The effect of all these steps is that we have arrived at a program
in completed form, i.e.\ in which all predicates are defined by
a single clause of the form $p(x_1, \ldots, x_n) \colondash G$,
where the free variables of $G$ are among $x_1, \ldots, x_n$.

Given program $P$, we refer to the program resulting at the end
of the sequence of transformations as the {\it augmented Clark
completion} of $P$, or $acc(P)$.
The augmented Clark completion of $P$ serves
essentially the same purpose as the Clark completion in Clark's
original treatment of negation as failure \cite{clark-negfail};
that is, it gives a closed form of the intended meaning of
each predicate.  We cannot truly consider it to be a logical
completion, however; the $if$ construct, while it can be
given a semantics consistent with the witness properties (as we
will see), cannot be given a logical interpretation.

\subsubsection{Example}
\label{example-subsection}

As an example, consider the first ``delete'' program from Section
\ref{example-section}:
\begin{tabbing}
\Indent{} \= \kill
\>  $d(x, [~], [~])$ \\
\>  $d(x, [x|ys], zs) \colondash !, d(x, ys, zs)$ \\
\>  $d(x, [y|ys], [y|zs]) \colondash d(x, ys, zs)$
\end{tabbing}
Assume that the variables selected in Step 1 are $x_1, x_2, x_3, \ldots$.
The program is transformed, by the end of Step 2, to the form:
\begin{tabbing}
\Indent{} \= \kill
\>  $d(x_1, x_2, x_3) \colondash (x_2=[~]), (x_3=[~])$ \\
\>  $d(x_1, x_2, x_3) \colondash (x_2=[x_1|ys]), !, d(x_1, ys, x_3)$ \\
\>  $d(x_1, x_2, x_3) \colondash (x_2=[y|ys]), (x_3=[y|zs]), d(x_1, ys, zs)$
\end{tabbing}
By the end of step 7, the program has been transformed into:
\begin{tabbing}
\Indent{} \= \kill
\>  $d(x_1, x_2, x_3) \colondash (x_2=[~] ~\And ~ x_3=[~])$ \\
\>  $d(x_1, x_2, x_3) \colondash (x_2=[x_1|ys]), !, d(x_1, ys, x_3)$ \\
\>  $d(x_1, x_2, x_3) \colondash (x_2=[y|ys] ~\And ~ x_3=[y|zs] ~\And ~ d(x_1, ys, zs))$
\end{tabbing}
Step 8 has no effect because there is no clause with more than
one cut (this is the case in most programs).  However, Step 9
scopes the local variables in the last clause, making the whole
program read as follows:
\begin{tabbing}
\Indent{} \= \kill
\>  $d(x_1, x_2, x_3) \colondash (x_2=[~] ~\And ~ x_3=[~])$ \\
\>  $d(x_1, x_2, x_3) \colondash (x_2=[x_1|ys]), !, d(x_1, ys, x_3)$ \\
\>  $d(x_1, x_2, x_3) \colondash \exists y \exists ys \exists zs(x_2=[y|ys] ~\And ~ x_3=[y|zs] ~\And ~ d(x_1, ys, zs))$
\end{tabbing}
Let us refer to this new body of the third clause as $B_3$.
Step 10 first combines the last two clauses into a single clause
with $if$, resulting in a new program as follows:
\begin{tabbing}
\Indent{} \= \Indent{} \= \Indent{} \= \kill
\>  $d(x_1, x_2, x_3) \colondash (x_2=[~] ~\And ~ x_3=[~])$ \\
\>  $d(x_1, x_2, x_3) \colondash$ \\
\>  \> $if[ys](x_2=[x_1|ys], d(x_1, ys, x_3)) \Or
                              (\Not\exists ys(x_2=[x_1|ys]) ~\And ~ B_3)$ \\
\end{tabbing}
Let us refer to this new body of the second clause as $B_2$.  Step
10 then continues, and transforms the remaining two clauses to the
single clause
\begin{tabbing}
\Indent{} \= \kill
\>  $d(x_1, x_2, x_3) \colondash (x_2=[~] ~\And ~ x_3=[~]) \Or B_2$
\end{tabbing}
The program is now in completed form.

\subsubsection{Properties}
\label{properties-subsection}

We now prove the properties we want the algorithm to have:
that is, that it terminates, that it produces a program in
completed form, and that the completed-form result program
actually does the same thing as the original program.

\begin{thm}[Completion Algorithm Termination]
  The completion algorithm terminates.
\end{thm}

\begin{proof}
Each loop in the algorithm continues while there is a
clause in the program with a specified property.  The effect of
each loop, however, is to eliminate all clauses with the
specified property.  Therefore each loop in the algorithm
terminates. 
\end{proof}

\begin{thm}[Completed Form Formation]
  The completion algorithm produces a program in completed form.
\end{thm}

\begin{proof}
Once the program being transformed achieves each of the
properties (A)-(E), as stated in the algorithm text, it never
loses those properties.  The conjunction of the properties
(A)-(E) is the same as saying that the program is in completed
form.
\end{proof}

To prove that the completion algorithm preserves the results of
computations, it is technically necessary to prove by induction
on the structure of computations that each transformation step
preserves result.  For brevity, we will prove this in detail
for only one of the transformations, and then argue more informally
in the main proof.  The following is a lemma and a theorem to do
with the transformation we will prove in detail.  All proofs are
contained in Appendix \ref{proofs-appx}.

\begin{lemma}
\label{completed-goal-stack-lemma}
  Let $\alpha$ be a goal stack.
  Let $\alpha'$ be $\alpha$
  with any number of occurrences of a sequence $B, C$ in a goal stack or
  clause body replaced by $B \And C$, where $B$ and $C$ are
  formulas.
  Then $(\theta: \alpha \Goes{P} \rho)$ in the liberal general semantics
  iff  $(\theta: \alpha' \Goes{P} \rho)$ in the liberal general semantics.
\end{lemma}

\begin{proof}
See Appendix \ref{proofs-appx}.  
\end{proof}

\begin{lemma}
\label{conjoin-lemma}
  Let $P'$ be $P$ with some sequence $B, C$ in a clause body
  replaced by $B \And C$.
  Then $\theta: \alpha \Goes{P} \rho$ in the liberal general semantics
  iff  $\theta: \alpha \Goes{P'} \rho$ in the liberal general semantics.
\end{lemma}

\begin{proof}
See Appendix \ref{proofs-appx}.  
\end{proof}

The main result preservation theorem is as follows.

\begin{thm}[Result Preservation of Completion Algorithm]
\label{result-preservation-thm}
  The completion algorithm preserves result according to
  the liberal general operational semantics.  That is,
  if $P'$ is the completion of $P$, then
  $\theta: \alpha \Goes{P} \rho$ in the liberal general semantics iff
  $\theta: \alpha \Goes{P'} \rho$ in the liberal general semantics.
\end{thm}

\begin{proof}
We prove the theorem by proving that each of the
transformations preserves result.  The lemma is used in the
proof of step 3.  The details of the proof can be found in
Appendix \ref{proofs-appx}.  
\end{proof}

Now that we know that the completion process
preserves result, we can assume that the programs we deal with
will be in completed form (since if not, we have an automatic
process for transforming them to completed form).  We will
therefore assume this for the rest of this paper.

\subsection{The Liberal Completed Semantics}
\label{libcomp-subsection}

Due to the complex behaviour of the Prolog cut, the liberal
general operational semantics contains nine rules for
predicates.  These rules exist mainly to manipulate the
sequences of body elements that exist in the clauses of a
general program, and to backtrack over multiple clauses defining
a predicate.  Since we now are assuming completed-form programs,
we can discard these rules in favour of one simple rule.  The
resulting operational semantics is referred to as the
{\it liberal completed} semantics.  Its simplicity moves us to
adopt it as the standard presentation of the liberal semantics
for the rest of the paper.

\begin{figure}[tp]

\noindent
\Rule{Pred:}
  {\theta: B[x_1:=t_1,\ldots,x_n:=t_n], \alpha \Goes{} \rho}
  {\theta: p(t_1, \ldots, t_n),         \alpha \Goes{} \rho}
  {where $p(x_1, \ldots, x_n) \colondash B$ is the clause
   defining $p$ in the completed-form program $P$}

\caption{
  The predicate rule for the liberal completed semantics,
  the only rule in the [Completed Predicates] fragment.
}
\label{completed-preds-rule-figure}
\end{figure}

The liberal general semantics' nine rules for predicates were
contained in the fragments [General Predicates].  The one rule
replacing them is the rule contained in Figure
\ref{completed-preds-rule-figure}.  We refer to the proof system
fragment containing only this rule as the [Completed Predicates]
fragment.  Thus, the liberal completed semantics consists of the
fragments [Basic], [Liberal Choice], and [Completed Predicates].

The following result proves that it is safe to use the
liberal completed semantics when we have a completed program.

\begin{thm}[Equivalence of General and Completed Semantics]
  If $P$ is a program in completed form, then the liberal
  general and liberal completed semantics have the same
  result.  That is,
  $\theta: \alpha \Goes{P} \rho$ in the liberal general
  semantics iff
  $\theta: \alpha \Goes{P} \rho$ in the liberal completed
  semantics.
\end{thm}

\begin{proof}
The computation in the liberal general semantics may
have portions ending in applications of the Using/nocut/succ and
Pred rules, of the following form.
\begin{center}
\begin{tabular}{c}
$\underline{\theta\xi: G\xi, \alpha\xi \Goes{} \rho}$ \\
$\vdots$ \\
$\overline{\theta: \vec{t} = \vec{x}, G, \alpha \Goes{} \rho}$ \\
$\overline{\underline{\theta: p(\vec{t}) using (p(\vec{x}) \colondash G), \alpha \Goes{} \rho}}$ \\
${\theta: p(\vec{t}), \alpha \Goes{} \rho}$ \\
\end{tabular}
\end{center}
where $\xi$ is the substitution $[x_1:=t_n, \ldots, x_n:=t_n]$.
(We assume without loss of generality that the free variables
of the clause are distinct from those of the conclusion.)
This portion of the computation in the liberal completed
semantics will have the following form:
\begin{center}
\begin{tabular}{c}
$\theta: G\xi, \alpha \Goes{} \rho'$ \\
$\overline{\theta: p(\vec{t}), \alpha \Goes{} \rho'}$ \\
\end{tabular}
\end{center}
where $\rho'$ differs from $\rho$ only in that it does not contain
substitutions for the renamed variables arising from clauses.
Since the substitution $\xi$ deals only with the $x_i$ variables,
which do not appear in $\alpha$, the uppermost judgements in the
two computations are essentially identical.

The computation in the liberal general semantics may also
have portions ending in a sequence of applications of the Using/empty,
Using/nocut/fail and Pred rules, of the following form.
\begin{center}
\begin{tabular}{ccc}
$\theta\xi: G\xi, \alpha\xi \Goes{} fail$ \\
\cline{1-1}
$\vdots$ \\
\cline{1-1} \cline{3-3}
$\theta: \vec{t} = \vec{x}, G, \alpha \Goes{} fail$ & \Sep &
$\theta: p(\vec{t}) using (), \alpha \Goes{} fail$ \\
\cline{1-3}
\multicolumn{3}{c}{
  $\theta: p(\vec{t}) using (p(\vec{x}) \colondash G), \alpha \Goes{} fail$
} \\
\cline{1-3}
\multicolumn{3}{c}{
  $\theta: p(\vec{t}), \alpha \Goes{} fail$
} \\
\end{tabular}
\end{center}
where $\xi$ is the substitution $[x_1:=t_n, \ldots, x_n:=t_n]$.
This portion of the computation in the liberal completed semantics
will have the following form:
\begin{center}
\begin{tabular}{c}
$\theta: G\xi, \alpha \Goes{} fail$ \\
$\overline{\theta: p(\vec{t}), \alpha \Goes{} fail}$ \\
\end{tabular}
\end{center}
Again, the substitution $\xi$ does not affect $\alpha$.
\end{proof}

\begin{figure}[tp]

\begin{center} 
\begin{tabular}{ccc}
& \Sep &
  $\overline{\underline{\theta'': \epsilon \Goes{} \theta''}}$ \\
& \Sep &
  $[ys:=[~]]: z=[~] \Goes{} \theta''$ \\
& \Sep &
  $\overline{[ys:=[~]]: [~]=[~], z=[~] \Goes{} \theta''}$ \\
& \Sep &
  $\overline{[ys:=[~]]: ([~]=[~] \And z=[~]) \Goes{} \theta''}$ \\
\cline{3-3}
$[ys:=[~]]: \epsilon \Goes{} [ys:=[~]]$
& \Sep &
  $[ys:=[~]]: ([~]=[~] \And z=[~]) \Or B_2 \Or B_3 \Goes{} \theta''$ \\
\cline{1-1} \cline{3-3}
$(): [a]=[a|ys] \Goes{} [ys:=[~]]$
  & \Sep &
  $[ys:=[~]]: d(a, [~], z)) \Goes{} \theta''$ \\
\cline{1-3}
\multicolumn{3}{c}{
  $(): if[ys]([a]=[a|ys], d(a, ys, z)) \Goes{} \theta''$
} \\
\cline{1-3}
\multicolumn{3}{c}{
  $(): if[ys]([a]=[a|ys], d(a, ys, z)) \Or B_3 \Goes{} \theta''$
} \\
\end{tabular}
\end{center}

\begin{center} 
\begin{tabular}{ccc}
~
  & \Sep &
  (see above) \\
\cline{1-1} \cline{3-3}
$(): [a]=[~] \And z=[~] \Goes{} fail$
  & \Sep &
  $(): if[ys]([a]=[a|ys], d(a, ys, z)) \Or B_3 \Goes{} \theta''$ \\
\cline{1-3}
\multicolumn{3}{c}{
  $\underline{(): ([a]=[~] \And z=[~]) \Or B_2 \Or B_3 \Goes{} \theta''}$
} \\
\multicolumn{3}{c}{
  $(): d(a,[a], z) \Goes{} \theta''$
} \\
\end{tabular}
\end{center}

\caption{A sample computation in the liberal completed semantics.
  $B_1 \Or B_2 \Or B_3$ is the body of the clause defining $d$
  from the second program in Section 3, with parameters
  instantiated.  Not all substitutions are listed in full.
}
\label{delete-example-lc-fig}
\end{figure}

Figure \ref{delete-example-lc-fig} shows a sample computation in
the liberal completed semantics, using the second, one-clause
version of the delete program from Section 3.  ($\theta''$ is the
substitution $[ys:=[~],z:=[~]]$.)  Note that
although the number of steps is similar to that of the liberal
general computation, now the elements of a goal stack are simply
formulas.  This will simplify our analysis,
since we can focus on formulas rather than having to
deal with the interaction of formulas and sequences of clauses
with cuts.

\subsection{Inadequacy of Liberal Semantics}
\label{inadequate-section}

\begin{figure}[tp]

\begin{center}
\begin{tabular}{ccc}
\cline{1-1}
$[x:=0]: \epsilon \Goes{} [x:=0]$ \\
\cline{1-1} \cline{3-3}
$(): x=0 \Goes{} [x:=0]$
  & \Sep &
  $[x:=1]: \epsilon \Goes{} [x:=1]$ \\
\cline{1-1} \cline{3-3}
$(): \Not(x=0) \Goes{} fail$
  & \Sep &
  $(): x=1 \Goes{} [x:=1]$ \\
\cline{1-3}
\multicolumn{3}{c}{
  $(): \Not(\Not(x=0)), x=1 \Goes{} [x:=1]$
} \\
\multicolumn{3}{c}{
  $\overline{(): \Not(\Not(x=0)) \And x=1 \Goes{} [x:=1]}$
} \\
\end{tabular}
\end{center}
\vspace{2mm}

\begin{center}
\begin{tabular}{ccc}
$\overline{(): \epsilon \Goes{} ()}$ \\
$\overline{(): 0=0 \Goes{} ()}$ \\
\cline{1-1} \cline{3-3}
$(): \Not(0=0) \Goes{} fail$
  & \Sep &
  $(): 0=1 \Goes{} fail$ \\
\cline{1-3}
\multicolumn{3}{c}{
  $(): \Not(\Not(0=0)), 0=1 \Goes{} fail$
} \\
\multicolumn{3}{c}{
  $\overline{(): \Not(\Not(0=0)) \And 0=1 \Goes{} fail}$
} \\
\end{tabular}
\end{center}
\vspace{2mm}

\begin{center}
\begin{tabular}{ccc}
\cline{1-1} \cline{3-3}
$(): a=0 \Goes{} fail$
  & \Sep &
  $(): \epsilon \Goes{} ()$ \\
\cline{1-3}
\multicolumn{3}{c}{
  $(): \Not(a=0) \Goes{} ()$
} \\
\multicolumn{3}{c}{
  $\overline{(): \Not(\Not(a=0)), a=1 \Goes{} fail}$
} \\
\multicolumn{3}{c}{
  $\overline{(): \Not(\Not(a=0)) \And a=1 \Goes{} fail}$
} \\
\end{tabular}
\end{center}

\caption{Computations showing that the goal 
  $\Not(\Not(x=0)) \And x=1$ violates the success property
  in the liberal completed semantics.
  $a$ is some arbitrary ground term not identical to 0.
}
\label{unsound-examplea-lc-fig}
\end{figure}

\begin{figure}[tp]

\begin{center}
\begin{tabular}{c}
\cline{1-1}
  $[x:=0]: \epsilon \Goes{} [x:=0]$ \\
\cline{1-1}
  $(): x=0 \Goes{} [x:=0]$ \\
\cline{1-1}
  $(): \Not(x=0), x=1 \Goes{} fail$ \\
\cline{1-1}
  $(): \Not(x=0) \And x=1 \Goes{} fail$ \\
\end{tabular}
\end{center}
\vspace{2mm}

\begin{center}
\begin{tabular}{ccc}
\cline{3-3}
~
  & \Sep &
  $(): \epsilon \Goes{} ()$ \\
\cline{1-1} \cline{3-3}
$(): 1=0 \Goes{} fail$
  & \Sep &
  $(): 1=1 \Goes{} ()$ \\
\cline{1-3}
\multicolumn{3}{c}{
  $(): \Not(1=0), 1=1 \Goes{} ()$
} \\
\multicolumn{3}{c}{
  $\overline{(): \Not(1=0) \And 1=1 \Goes{} ()}$
} \\
\end{tabular}
\end{center}

\caption{Computations showing that the goal 
  $\Not(x=0) \And x=1$ violates the failure property
  in the liberal completed semantics.
}
\label{unsound-exampleb-lc-fig}
\end{figure}

Because it is intended to capture the behaviour of Prolog
programs with cut, the liberal completed semantics does not
have either of the witness properties.  Figure
\ref{unsound-examplea-lc-fig} shows that the goal formula
$G_1 \equiv \Not(\Not(x=0)) \And x=1$ succeeds in the liberal
completed semantics, even though
$G_1[x:=0]$ fails and $G_1[x:=a]$, where $a$ is any arbitrary
ground term not identical to 0, fails.
Similarly, Figure \ref{unsound-exampleb-lc-fig} shows that the
goal formula $G_2 \equiv \Not(x=0) \And x=1$ fails in the liberal
completed semantics, even though
$G_2[x:=1]$ succeeds.

This is consistent with the behaviour of
the usual unsound implementation of negation as failure.  We
can, of course, ban unsound NAF alone with a mode restriction
similar to that of \Staerk{} \cite{staerk-lptp-jlp}; however, if
we retain the general {\it if} construct (corresponding to the hard
cut), we will still permit behaviour which violates the witness
properties.  This suggests that we need some further restriction
to {\it if} analogous to \Staerk{}'s restriction on negation.

Note that these counterexamples also show that the liberal
general semantics (a generalization of the liberal completed
semantics) has neither of the witness properties.

\section{The Conservative Operational Semantics}
\label{conservative-section}

In the last section, we gave operational semantics for programs
which characterized Prolog computation, but were inadequate from
a logic-programming point of view because they violated
the witness properties.  In this section, we repair the faults
of the liberal semantics by placing simple restrictions on some
of its rules.  The result is the {\it conservative} semantics,
which does enjoy the witness properties.  We refer to the form
of cut embodied in the conservative semantics as {\it firm} cut.

In Section \ref{conservative-rules-section}, we present and
describe the rules for the conservative semantics, and in
Section \ref{conservative-properties-section} we prove useful
properties of it, including the witness properties.  Finally, in
\ref{implementation-section} we show that the firm cut still
permits the useful first-solution behaviour of the Prolog cut.

\subsection{The Conservative Semantics Rules}
\label{conservative-rules-section}

\begin{figure}[tp]

\noindent
\Rule{Not/succ:}
  {\theta: B \Goes{} \theta'}
  {\theta: \Not B, \alpha \Goes{} fail}
  {where $B$ has no free variables}
\\ %
\Rule{Not/fail:}
  {\theta: B \Goes{} fail  \Sep  \theta: \alpha \Goes{} \rho}
  {\theta: \Not B, \alpha \Goes{} \rho}
  {where $B$ has no free variables}
\\ %
\Rule{Not/flounder:}
  {}
  {\theta: \Not B, \alpha \Goes{} flounder}
  {where $B$ has free variables}
\\ %
\Rule{Not/sub:}
  {\theta: B \Goes{} \rho}
  {\theta: \Not B, \alpha \Goes{} \rho}
  {where $B$ has no free variables, and $\rho$ is $flounder$ or $diverge$}
\\ %
\Rule{If/succ:}
  {\theta: B[\vec{x}:=\vec{x}'] \Goes{} \theta'  \Sep
   \theta': C[\vec{x}:=\vec{x}']\theta', \alpha \Goes{} \rho}
  {\theta: if[\vec{x}](B,C), \alpha \Goes{} \rho}
  {where $\exists \vec{x}(B)$ has no free variables, and
   $\vec{x}'$ do not appear in the conclusion}
\\ %
\Rule{If/fail:}
  {\theta: B[\vec{x}:=\vec{x}'] \Goes{} fail}
  {\theta: if[\vec{x}](B,C), \alpha \Goes{} fail}
  {where $\exists \vec{x}(B)$ has no free variables, and
   $\vec{x}'$ do not appear in the conclusion}
\\ %
\Rule{If/flounder:}
  {}
  {\theta: if[\vec{x}](B,C), \alpha \Goes{} flounder}
  {where $\exists \vec{x}(B)$ has free variables}
\\ %
\Rule{If/sub:}
  {\theta: B[\vec{x}:=\vec{x}'] \Goes{} \rho}
  {\theta: if[\vec{x}](B,C), \alpha \Goes{} \rho}
  {where $\exists \vec{x}(B)$ has no free variables, and
   $\vec{x}'$ do not appear in the conclusion, and $\rho$
   is $flounder$ or $diverge$}

\caption{
  The rules of the [Conservative Choice] fragment, for
  computing the choice constructs in a more restricted fashion.
}
\label{conservative-choice-fragment-fig}
\end{figure}

\begin{figure}[tp]

\begin{center}
\begin{tabular}{c}
\cline{1-1}
  $(): \Not(x=0), x=1 \Goes{} flounder$ \\
\cline{1-1}
  $(): \Not(x=0) \And x=1 \Goes{} flounder$ \\
\end{tabular}
\end{center}

\caption{The safe computation of
  $\Not(x=0) \And x=1$ in the conservative semantics.
}
\label{unsound-exampleb-c-fig}
\end{figure}

The conservative operational semantics 
restricts the computation of negation and {\it if}.
Whereas the liberal completed semantics is made up of the
rules fragments
[Basic] (Fig.\ \ref{basic-rules-fig}),
[Liberal Choice] (Fig.\ \ref{lib-choice-rules-fig}), and
[Completed Predicates] (Fig.\ \ref{completed-preds-rule-figure}),
the conservative semantics is made up of the rules fragments
[Basic], [Conservative Choice], and [Completed Predicates].
The rules for the new fragment, [Conservative Choice], are in
Figure \ref{conservative-choice-fragment-fig}.

Consider the rules Not/succ and Not/fail from [Conservative
Choice].  These rules are the same as those of the [Liberal
Choice] fragment, except that they have the restriction that $B$
(the negated formula) must have no free variables.  When $B$
does have free variables, a new rule, Not/flounder, applies.
Not/flounder states that a goal stack beginning with a negated
formula with free variables immediately returns a new result,
$flounder$, indicating that the computation cannot continue
at this point.

Another new rule, Not/sub, defines what happens when $B$ has no
free variables, but the sub-computation itself flounders: the
$flounder$ result is passed on.  Note that the rules in the
[Basic] and [Completed Predicates] fragments are already described in
such a way that they also automatically pass on the new $flounder$
result.  Hence, $flounder$ acts as a kind of run-time exception,
which causes the computation to terminate immediately.\footnote{
  Not/sub also passes on the result $diverge$, which is not
  needed until Section \ref{pessimistic-sec}.
}

The conservative rules for the $if$ connective are constructed
from those of the liberal rules in a similar manner, modulo the
bound variables of the $if$.  As an example of a conservative
computation, consider again the goal $\Not(x=0) \And x=1$,
which was a problem for the liberal semantics.  Figure
\ref{unsound-exampleb-c-fig} shows that the conservative
semantics handles it in a sound way, by immediately stating that
it flounders.

We should note at this point that there are other approaches
to the problem of handling negation in a sound way.  Loveland and Reed,
for example \cite{loveland-near-horn}, define a resolution
method by which queries against programs with negation
can be evaluated in a sound and complete manner.  Dahl
\cite{dahl-negation} defines an approach which delays the
evaluation of a negated goal until it becomes ground, and
an approach which, within a negated goal's computation,
blocks only the unification of variables which are free
outside the scope of the negation.
Di Pierro et al.\ \cite{IC::PierroMP1995}
define an approach in which an existentially closed negated atom
(a formula of the form $\exists[\Not A]$)
succeeds iff all branches of the SLD-tree of the atom either
fail or instantiate the atom.  Some of these methods
have been implemented in a variety of systems, for instance in
Naish's NU-Prolog \cite{naish-negation-lncs}.  Here we are
motivated by our interest in the features implemented
in the most widely-used Prolog systems.  Most Prolog systems
implement the simple negation as failure characterized by
the liberal semantics and restricted by the conservative
semantics.

\subsection{Properties of the Conservative Semantics}
\label{conservative-properties-section}

In this section, we prove the properties of the conservative
operational semantics that we wanted to hold.  First, we prove
the correspondence of computations in the liberal and the
conservative semantics.  Then, we prove the witness properties.

\subsubsection{Correspondence of Computations}

First, we note that successful and failing computations in
the conservative semantics correspond to successful and
failing computations in the liberal completed semantics.

\begin{thm}
  If $\theta: \alpha \Goes{} \rho$ in the conservative
  semantics, and $\rho$ is not $flounder$, then
  $\theta: \alpha \Goes{} \rho$ in the liberal completed
  semantics.
\end{thm}

\begin{proof}
Any computation in the conservative semantics which
contains applications of the Not/flounder or If/flounder rules
must result in $flounder$, since the $flounder$ outcomes of
these rules descend through all other rules in the [Basic],
[Conservative choice] and [Completed predicates] fragments.
Therefore if a computation in the conservative semantics
does not result in $flounder$, it must not use those rules;
rather, it uses only the
other Not and If rules, which are
restrictions of those in the liberal completed semantics,
and the other rules, which are identical to those in the
liberal completed semantics.
Such a computation is, in fact, a computation in the liberal
completed semantics.  
\end{proof}

The converse does not hold, since successful and failing
computations in the liberal completed semantics may flounder
in the conservative semantics.  However, all
computations in the liberal completed semantics do correspond
to some kind of computations in the conservative semantics,
as the next theorem shows.

\begin{thm}
  If $\theta: \alpha \Goes{} \rho$ in the liberal completed
  semantics, then there is some $\rho'$ such that
  $\theta: \alpha \Goes{} \rho'$ in the conservative
  semantics, and $\rho'$ is either $\rho$ or $flounder$.
\end{thm}

\begin{proof}
By induction on the structure of the liberal completed
computation.  Cases are on the bottommost rule application.

All applications of rules with 0 premises correspond to rule
applications in the conservative semantics.

If the bottommost rule is Disj/fail:
the bottommost judgement is of the form
$(\theta: B \Or C, \alpha \Goes{} fail)$,
and its left-hand premise judgement
is of the form $(\theta: B, \alpha \Goes{} fail)$.
By the induction hypothesis (IH), either 
$(\theta: B, \alpha \Goes{} fail)$ in the conservative semantics,
or $(\theta: B, \alpha \Goes{} flounder)$ in the conservative semantics.
In the first case, the result follows directly from another
application of the IH;
in the second case, the result follows from one application of the
Disj/nofail rule.

The cases for the Not and If rules are similar to that of Disj/fail.
Applications of all other
rules in the liberal completed computation have exactly one
premise, and correspond to applications of the same rules in the
conservative computation. 
\end{proof}

Examples of goals whose outcomes differ in the liberal completed
and conservative semantics are as follows:
\begin{itemize}
\item The goal $\Not\Not(x=0)$ succeeds in the liberal
  completed semantics, but flounders in the conservative semantics.
\item The goal $\Not(x=0)$ fails in the liberal
  completed semantics, but flounders in the conservative semantics.
\item The goal $\Not\Not(x=0) \And loop(x)$,
  where the predicate $loop$ is defined with the definition
  $loop(x) \colondash loop(x)$, diverges (does not have any
  finite computation) with respect to the liberal completed
  semantics; however, it flounders in the conservative
  semantics.
\end{itemize}
These examples, along with the witness properties to be proven
next, show that although strictly fewer goals succeed or fail in
the conservative semantics, strictly more goals terminate in the
conservative semantics.

\subsubsection{The Witness Properties}
\label{witness-properties-section}

Finally, we show the witness properties of the conservative
semantics.  Most proofs are contained in full in Appendix
\ref{proofs-appx}.

We begin with some useful definitions.
We say that {\it $\theta$ is a specialization of $\theta'$},
in symbols $\theta \subseteq \theta'$, if there is some
$\theta''$ such that $x\theta \equiv x\theta'\theta''$,
for all variables $x$ in the domain of $\theta'$.
Given a set $V$ of variables and a substitution $\theta$,
we say that a substitution $\xi$ {\it grounds $V$ consistent
with $\theta$} if $\xi \subseteq \theta$ and $x\xi$ is
ground for every $x \in V$.

An inductive generalization of the failure property can be
proven directly; the corresponding generalization of the
success property requires a technical lemma.  These three
lemmas are as follows.

\begin{lemma}[General Failure Property of Conservative Semantics]
  \label{general-failure-property-lemma}
  Let $\theta, \alpha$ be such that
  $(\theta: \alpha \Goes{} fail)$ in the conservative semantics.
  Then for any $\xi$,
  $(\theta: \alpha\xi \Goes{} fail)$ in the conservative semantics.
\end{lemma}

\begin{proof}
See Appendix \ref{proofs-appx}. 
\end{proof}

\begin{lemma}[Substitution Monotonicity of Conservative Semantics]
  \label{substitution-monotonicity-lemma}
  Let $\theta, \alpha$ be such that $\alpha\theta \equiv \alpha$ and
  $\theta: \alpha \Goes{} \theta'$ in the conservative semantics.
  Then $\theta' \subseteq \theta$.
\end{lemma}

\begin{proof}
See Appendix \ref{proofs-appx}. 
\end{proof}

\begin{lemma}[General Success Property of Conservative Semantics]
  \label{general-success-property-lemma}
  Let $\theta, \alpha$ be such that
  $\theta: \alpha \Goes{} \theta'$ in the conservative semantics.
  Let $V$ be a subset of the free variables of $\alpha$.
  Then for any $\xi$ grounding $V$ consistent with $\theta'$,
  $\theta: \alpha\xi \Goes{} \theta'\xi$ in the conservative semantics.
\end{lemma}

\begin{proof}
See Appendix \ref{proofs-appx}. 
\end{proof}

We can now state and prove the witness properties mentioned in
the Introduction for the conservative semantics.  First, we
define more precisely what we mean by success and failure.

We say that a goal $G$ {\it succeeds} (in the conservative semantics)
if there is a computation with a conclusion of the form $(): G \Goes{} \theta'$.
We say that a goal $G$ {\it fails}
if there is a computation with a conclusion of the form $(): G \Goes{} fail$.

\begin{thm}[Witness Properties of the Conservative Semantics] ~ \\
  (1) If a goal $G$ succeeds, then some ground instance of $G$ succeeds. \\
  (2) If a goal $G$ fails, then any ground instance of $G$ fails.
\end{thm}

\begin{proof}
(1) If $G$ succeeds, this means there is a $\theta'$ such that
$(): G \Goes{} \theta'$.  Let $\sigma$ be the substitution which substitutes
all free variables of $G\theta'$ by 0.  Let $\xi$ be the substitution
which substitutes any variable $x \in FV(G)$ by $x\theta'\sigma$.
Then $\xi$ grounds $FV(G)$ consistent with $\theta'$. By the
General Success Property, we have that $(): G\xi \Goes{} \theta'\xi$.
Thus the ground instance $G\xi$ of $G$ succeeds.

(2) If $G$ fails, then $(): G \Goes{} fail$.  By the General Failure
Property, for any $\xi$, including those grounding all variables
in $FV(G)$, we have that $(): G\xi \Goes{} fail$.  Thus all ground
instances of $G$ fail.  
\end{proof}

\subsection{Implementation Issues}
\label{implementation-section}

In this section, we discuss some implementation-related issues.
We show that the conservative semantics retains the
desirable first-solution behaviour of the Prolog hard cut.
We also discuss the possibility of turning the mode restriction
of the conservative semantics into a static rather than a
dynamic one.

\subsubsection{First Solution Behaviour}

When we have a formula of the form $if[\vec{x}](B, C)$,
the conservative operational semantics allows the $\vec{x}$
variables to pass on to $C$, and allows free variables other
than $\vec{x}$ in $C$; however, only the first successful
substitution for $\vec{x}$ is passed on.  The conservative
semantics therefore still allows the useful ``first
solution'' behaviour which $if$ has inherited from cut.

For an example of this behaviour, consider the following
problem.  We define an {\em association list} as a list of terms
of the form $a(k,j)$, where $k$ is a key and $j$ is a value
associated with it.  A problem commonly encountered in symbolic
programming is to extract the first value (and only the first
value) associated with a key in an association list, which is
taken as the ``current'' value of the key.
We can write the standard logic programming ``member'' predicate as
\[ m(x,y) \colondash
   \exists yh \exists yt (y=[yh|yt] \And (x=yh \Or m(x,yt))) \]
and then write a predicate which solves the first-value problem as follows:
\[ v(x, y, z) \colondash if[w](m(a(y,w), x), z=w) \]
The predicate call $v(x, y, z)$, where $x$ is an association
list, $y$ is a key, and $z$ is any term, succeeds iff $z$ is the
first value associated with $y$ in $x$.

\begin{figure}

\noindent
Computation of $m$ subgoal:
\begin{center}
\begin{tabular}{c}
$\overline{
  [yh:=a(b,0),yt:=[a(b,1)],w:=0]: \epsilon \Goes{} [w:=0]
}$ \\
$\overline{
  [yh:=a(b,0),yt:=[a(b,1)]]: a(b,0)=a(b,w) \Goes{} [w:=0]
}$ \\
$\overline{\underline{
  [yh:=a(b,0),yt:=[a(b,1)]]: a(b,0)=a(b,w) \Or m(a(b,w),[a(b,1)] \Goes{} [w:=0]
}}$ \\
$(): [a(b,0),a(b,1)]=[yh|yt], (yh=a(b,w) \Or m(a(b,w),yt) \Goes{} [w:=0]
$ \\
$\overline{
  (): [a(b,0),a(b,1)]=[yh|yt] \And (yh=a(b,w) \Or m(a(b,w),yt) \Goes{} [w:=0]
}$ \\
$\overline{
  (): \exists yt([a(b,0),a(b,1)]=[yh|yt] \And (yh=a(b,w) \Or m(a(b,w),yt)))
      \Goes{} [w:=0]
}$ \\
$\overline{\underline{
  (): \exists yh \exists yt([a(b,0),a(b,1)]=[yh|yt] \And (yh=a(b,w) \Or m(a(b,w),yt)))
      \Goes{} [w:=0]
}}$ \\
$(): m(a(b,w),[a(b,0),a(b,1)]) \Goes{} [w:=0]$ \\
\end{tabular}
\end{center}

\noindent
Successful computation:
\begin{center}
\begin{tabular}{ccc}
\cline{3-3}
(see above) & & $[w:=0,z:=0]: \epsilon \Goes{} [z:=0]$ \\
\cline{1-1} \cline{3-3}
$(): m(a(b,w),[a(b,0),a(b,1)]) \Goes{} [w:=0]$
  & \Sep &  $[w:=0]: z=0 \Goes{} [z:=0]$ \\
\cline{1-3}
\multicolumn{3}{c}{
  $\underline{(): if[w](m(a(b,w),[a(b,0),a(b,1)]), z=w) \Goes{} [z:=0]}$
} \\
\multicolumn{3}{c}{
  $(): v([a(b,0),a(b,1)], b, z) \Goes{} [z:=0]$
} \\
\end{tabular}
\end{center}

\noindent
Failing computation:
\begin{center}
\begin{tabular}{ccc}
(see above) \\
\cline{1-1} \cline{3-3}
$(): m(a(b,w),[a(b,0),a(b,1)]) \Goes{} [w:=0]$
  & \Sep &  $[w:=0]: 1=0 \Goes{} fail$ \\
\cline{1-3}
\multicolumn{3}{c}{
  $\underline{(): if[w](m(a(b,w),[a(b,0),a(b,1)]), 1=w) \Goes{} fail}$
} \\
\multicolumn{3}{c}{
  $(): v([a(b,0),a(b,1)], b, 1) \Goes{} fail$
} \\
\end{tabular}
\end{center}

\caption{
  Examples showing first-solution behaviour of conservative
  semantics.  (Some substitutions are simplified for clarity.)
  Top: a computation returning the first solution to a call to the
  membership predicate.  Middle:  a computation
  showing that the first solution is selected by $if$.
  Bottom: a computation showing that subsequent solutions are not
  selected by $if$.
}
\label{alist-example-revised-fig}
\end{figure}

The query $v([a(b,0),a(b,1)], b, z)$ to this program should
result in the binding $[z:=0]$, since this is the first value
returned by $m$ as associated with the key $b$ in the list.
However, the query $v([a(b,0),a(b,1)], b, 1)$ to this program
should fail; even though the value 1 is associated with $b$
later in the list, $if$ should select only the first solution.
Figure \ref{alist-example-revised-fig} shows that this is indeed
the behaviour of the conservative semantics.

We could evidently get closer to the liberal general
semantics by allowing the first subformula of the $if$
to be computed with free variables,
as long as those variables do not get bound in the course of
the computation, as suggested by one of Dahl's negation
strategies \cite{dahl-negation} and Di Pierro et al.\ \cite{IC::PierroMP1995}.
Since this
would complicate the operational semantics and our analysis,
we have decided to stick with the conservative
semantics as given.

\subsubsection{Static Analysis}

\label{mode-checking-discussion-section}
The conservative operational semantics restricts the behaviour
of the logic programming system by essentially enforcing mode
checks at run time.  However, we do not believe that there is
any obstacle to doing static mode checking (see for example
\cite{barbuti-martelli-1990,apt-marchiori,gabbrielli-etalle-1999})
in order to catch programs at compile time which
could result in floundering goals.  (In \cite{andrews-cut-tr},
a static analysis scheme is proposed which does a
fine-grained analysis in order to reject as few programs as
possible, at the expense of some complexity.)


Because the conservative semantics behaves identically to the
liberal semantics on non-floundering goals, and because the
liberal semantics characterizes Prolog, we believe that an
implementation of firm cut is achievable simply by
imposing static mode restrictions on a conventional logic
programming system.  For the sake of brevity, we do not explore
this issue further here, but assume in the rest of the paper
that such a static analysis system is possible.

\section{The Abstract Semantics}
\label{sems-section}

In this section, we present an abstract semantics for the
conservative operational semantics.  The abstract semantics does
not reify such notions as substitution sequence and unification;
rather, the central element of the semantics which deals with
free variables is the interpretation of the existential quantifier
by a valuation function of the same form as those of classical
truth theory \cite{kripke,fitting-kripke}.
This suggests that the conservative semantics and firm cut
have a deeper connection to logic than simply permitting some
logical computations.

The abstract semantics is in the UNV
(unfolding-normal-form-valuation) style \cite{andrews-lnaf-tcs},
and it depends on the witness properties to achieve soundness
and completeness.  In UNV semantics, we associate a truth value
to a goal; the truth value can be described as the maximally
defined truth value among the valuations of the normal forms of
the unfoldings of the goal.  We doubt that it is possible to
give such a semantics for the liberal semantics and thus for
Prolog with hard cut, due to those systems' failure to achieve
the witness properties.

We begin with an overview of UNV semantics in
Section \ref{overview-section}
containing some basic definitions, including that of an
(operational) {\it outcome} of a goal $G$ with respect to a
program $P$, $outcome_P(G)$.  Section \ref{overview-section}
also contains a ``roadmap'' of the series of results that
follow, referred to as the ``raising lemmas''.
In Sections \ref{unfoldings-section} through \ref{denotation-section}
we proceed, through the raising lemmas, to systematically raise the
characterizing expression for $outcome_P(G)$ to greater and
greater levels of abstraction, until all operational notions
have been abstracted away.

Finally, in Section \ref{denotation-section}, we link the
previous raising lemmas into a final characterization of outcome
of a general goal with respect to a program, and give an
expression describing the abstract denotation of a program.
We conclude with an example, in Section
\ref{sems-example-section}, and some discussion in Section
\ref{sems-discussion-section}.

In this section, whenever we refer to a program $P$ and a goal
$G$, we assume that $G$ does not yield the $flounder$ result.
It may also be possible to characterize the $flounder$ result,
as in, for instance, \cite{andrews-lnaf-tcs}.
However, for simplicity, here we assume that programs will be subject
to a static analysis which excludes those able to generate such
a result, as discussed in Section \ref{mode-checking-discussion-section}.

\subsection{UNV Semantics}
\label{overview-section}

\begin{figure}[tp]

\centerline
{\epsffile{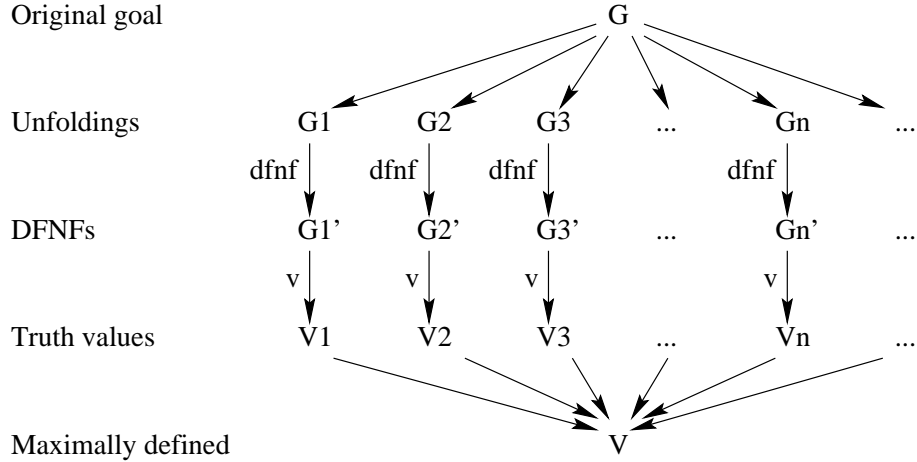}
}\vspace{5mm}

\caption{
  Diagram of the basic notions of UNV
  (unfolding-normal-form-valuation) semantics.
}
\label{unv-fig}
\end{figure}

Here we give an overview of UNV semantics and some basic
definitions which will be used throughout the section.
We also give a ``roadmap'' of the results which will be
proven.

\subsubsection{Overview}

The UNV semantics given here is based on six basic notions:
\begin{itemize}
\item The three {\it truth values} $T$, $F$ and $U$, or
  ``true'', ``false'', and ``undefined''.
\item The {\it definedness ordering} on truth values, which ranks
  $T$ and $F$ as being more defined than $U$.
\item The {\it alethic} or {\it truth ordering} on truth values,
  which ranks $U$ as ``more true'' than $F$ and $T$ as ``more
  true'' than $U$.
\item The {\it unfoldings} of a goal, which are the formulas
  obtained from the goal by expanding zero or more predicate
  calls, possibly repeatedly.
\item The {\it depth-first normal form}, or DFNF, of a goal,
  which is a formula closely related to the disjunctive
  normal form (DNF) of the goal.
\item The {\it valuation} $v(G)$ of a goal $G$ in DFNF, which is
  a compositional function from formulas to truth values.
\end{itemize}
The last three of these will be given more precise and detailed
definitions in the course of this section.

A schematic diagram of the basic notions of UNV semantics is
contained in Figure \ref{unv-fig}.  Given a goal $G$, we
consider all the (possibly infinitely many) unfoldings
$G_1, G_2, G_3, \ldots, G_n, \ldots$
of the goal.  Then, we find the DFNFs of all the unfoldings,
resulting in the normal-form goals
$G'_1, G'_2, G'_3, \ldots, G'_n, \ldots$.
We apply the valuation function
$v$ to the normal-form goals, getting a set
$V_1, V_2, V_3, \ldots, V_n, \ldots$ of truth values, each of
them equal to either $T$, $F$, or $U$.
(The alethic ordering of truth values is used to compute the
valuation of existentially-quantified goals.)
There will be one unique maximally defined truth value in this
set; this will be taken as the truth value of the original goal $G$.

\subsubsection{Outcomes of Goals}

When we evaluate a goal in a logic programming system, we expect
to receive a substitution (if one exists) as the result of the
evaluation.  However, when we prove properties of logic programs,
we are more interested in proving whether a general pattern of
goals succeeds or fails; we are less interested in obtaining
substitutions, because there may be a different substitution for
each different instance of the pattern.  Hence, in this paper (as in
\cite{andrews-phd-dd,andrews-lnaf-tcs,staerk-lptp-jlp}) we take
the ``observable'' of interest to be whether a goal succeeds, fails
or diverges, linking these observables to the truth values $T$,
$F$ and $U$ respectively.

We therefore define the {\it outcome} of a goal $G$ with
respect to $P$, $outcome_P(G)$, as follows.
\begin{itemize}
\item If there is a $\theta'$ such that $((): G \Goes{P} \theta')$
  in the conservative operational semantics, then $outcome_P(G) = T$.
\item If $((): G \Goes{P} fail)$ in the conservative operational semantics,
  then $outcome_P(G) = F$.
\item Otherwise (i.e., if there is no result $\rho$ such that
  $((): G \Goes{P} \rho)$ in the conservative semantics), then
  $outcome_P(G) = U$.
\end{itemize}
This notion of outcome will be what is characterized by the
abstract, UNV semantics.

For use in the raising lemmas, we will also need the
closely-related notion of ``pessimistic outcome''
$outcome^\frown(G)$ of a goal $G$.  This is what the
outcome of $G$ would be, independent of the program, if we were
to pessimistically assume that all predicates in the program
would diverge (result in infinite computations).  This notion
will be defined more precisely below.

\subsubsection{Roadmap}

Here we present a guide to the characterization results that
follow.  The sequence
of raising lemmas we will prove will be as follows:
\begin{enumerate}
\item The outcome of a goal $G$ with respect to a program $P$
  can be obtained by inspecting all the pessimistic outcomes of
  all the unfoldings of $G$, and taking the maximally defined one.
  ($outcome_P(G) = max_k(\{outcome^\frown(G') ~|~ G'$ is an
  $P$-unfolding of $G\})$.)
\item The pessimistic outcome of a goal $G$ is the same as the
  pessimistic outcome of its depth-first normal form.
  ($outcome^\frown(G) = outcome^\frown(dfnf(G))$.)
\item The pessimistic outcome of a goal $G$ in depth-first normal
  form can be characterized by a compositional valuation function
  (function from goals to truth values), $v$.
  ($outcome^\frown(G) = v(G)$.)
\item Putting the previous three raising lemmas together,
  the outcome of $G$ with respect to $P$, $outcome_P(G)$,
  can be alternatively characterized by the expression
  $max_k(\{v(dfnf(G')) ~|~ G'$ is a $P$-unfolding of $G\})$.
\end{enumerate}

This final result gives an abstract view of the meaning of a
program, which allows us to define the program's denotation,
concluding the characterization.

\subsection{Unfoldings and the Pessimistic Semantics}
\label{unfoldings-section}

In this section, we define the notion of unfolding of a goal,
and also define the pessimistic operational semantics, which treats all
predicates as being divergent.  We then show how the two notions
are related by proving that every terminating goal has some
unfolding which terminates even in the pessimistic semantics.
This property is useful because it allows us to abstract
away (into the notion of unfolding) all consideration
of the program, and concentrate on characterizing outcomes
under the program-independent pessimistic semantics.

We then draw upon the standard notion of definedness ordering
of truth values in order to get a succinct characterization
of this relationship.  The section concludes with the first
raising lemma.

\subsubsection{Unfoldings}

Informally, an unfolding of a goal is the goal after some
predicate calls are replaced by the corresponding predicate
bodies, possibly repeatedly.  The notion comes originally from
Burstall and Darlington's corresponding functional programming
notion \cite{burstall-darlington}, and is analogous to Tamaki
and Sato's notion of unfolding of a program
\cite{tamaki-sato-unfold}.  Unfoldings are also used in the
unfolding semantics of Gabbrieli and Levi \cite{levi-unfold-tcs92},
and in other semantics such as Etalle's for modular general
logic programs \cite{etalle-modular-general}.

More formally, given a program $P$ in completed form, a
formula $G'$ is a {\it 1-$P$-unfolding} of $G$ if it is $G$ with
one occurrence of $p(t_1, \ldots, t_n)$ replaced by
$B[x_1:=t_1, \ldots, x_n:=t_n]$, where
$(p(x_1, \ldots, x_n) \colondash B)$ is a definition in $P$.
A formula $G'$ is a {\it $P$-unfolding} of $G$ if it is either $G$
itself, or a $P$-unfolding of a 1-$P$-unfolding of $G$.
We will drop the program name $P$ when it is unimportant or
clear from context.  Clearly, the $P$-unfolding operation, seen
as a rewriting, is confluent.

For instance, let the program $P$ consist of the definitions
$(q \colondash r)$ and $(p \colondash q \And p)$.
Then the goal $G = (q \Or p)$ has two 1-$P$-unfoldings, namely
$(r \Or p)$ and $(q \Or (q \And p))$.  $G$ has an infinite number
of $P$-unfoldings, including $G$ itself, its two 1-$P$-unfoldings,
and other unfoldings such as $(r \Or (q \And (r \And p)))$.

We define a $P$-unfolding of a {\it sequence} $G_1, \ldots, G_n$
of formulas as any sequence $G'_1, \ldots, G'_n$ of formulae in
which $G'_i$ is a $P$-unfolding of $G_i$, for all $1 \leq i \leq n$.

\subsubsection{The Pessimistic Semantics}
\label{pessimistic-sec}

If we unfold a succeeding or failing goal enough, we obtain a
goal which succeeds or fails without doing any predicate
expansions.  A divergent goal, however, cannot be unfolded to
a point where it succeeds or fails without predicate expansions.

These facts suggest the following analytical framework.
We define an operational semantics, the {\it pessimistic} semantics,
which returns the result $diverge$ on any predicate
call.  We can then characterize a successful goal as one with an
unfolding which succeeds in the pessimistic semantics, a failing
goal as one with an unfolding which fails in the pessimistic
semantics, and a divergent goal as one with no unfolding which
returns anything but $diverge$ in the pessimistic semantics.

\begin{figure}[tp]

\noindent
\Rule{Pred:}
  {}
  {\theta: p(t_1, \ldots, t_n),         \alpha \Goes{} diverge}
  {}

\caption{
  The predicate rule for the pessimistic semantics,
  the only rule in the [Pessimistic Predicates] fragment.
}
\label{pessimistic-preds-rule-fig}
\end{figure}

To this end, we define the pessimistic operational semantics
as being made up of the the operational semantics fragments
[Basic], [Conservative Choice], and [Pessimistic Predicates],
where the latter fragment consists of the single rule shown in
Figure \ref{pessimistic-preds-rule-fig}.  Note that the rules
in [Basic] and [Conservative Choice] are described in such a way
that they pass on the $diverge$ outcome.  Thus, as soon as a
predicate call is encountered in the course of computation,
the pessimistic semantics effectively assumes that the
computation will diverge.  This means, for instance, that if there
is a predicate call in a goal $G$ to the left of the first
disjunction in $G$, then $G$ will diverge according to the
pessimistic semantics.

We define the {\it pessimistic outcome} of a goal $G$,
$outcome^\frown(G)$, as follows.
\begin{itemize}
\item If there is a $\theta'$ such that $((): G \Goes{} \theta')$
  in the pessimistic operational semantics, then $outcome^\frown(G) = T$.
\item If $((): G \Goes{} fail)$ in the pessimistic operational semantics,
  then $outcome^\frown(G) = F$.
\item Otherwise (i.e., if $((): G \Goes{} diverge)$ in the
  pessimistic semantics), then $outcome^\frown(G) = U$.
\end{itemize}
Note that the program $P$ is irrelevant to the pessimistic
semantics, and that all computations in the pessimistic
semantics are of bounded size because predicate calls are not
expanded.

\subsubsection{Results}

Here we show the relationship between unfoldings and the
pessimistic semantics.

\begin{thm}
\label{cons-unf-thm}
  Let $\theta: \alpha \Goes{P} \rho$ in the conservative semantics.
  Then some $P$-unfolding $\alpha'$ of $\alpha$ is such that
  $\theta: \alpha' \Goes{P} \rho$ in the pessimistic semantics.
\end{thm}

\begin{proof}
By induction on the structure of the conservative
computation.  Cases are on the bottommost rule, and all cases follow
trivially from the induction hypothesis except the case in which
the bottommost rule is a Pred rule.  In this case, one additional
predicate unfolding is necessary to obtain $\alpha'$ from the
$\alpha'$ of the induction hypothesis. 
\end{proof}

The converse of the above theorem is also the case:

\begin{thm}
\label{unf-cons-thm}
  Let some $P$-unfolding $\alpha'$ of $\alpha$ be such that
  $\theta: \alpha' \Goes{P} \rho$ in the pessimistic semantics,
  where $\rho$ is not $diverge$.
  Then $\theta: \alpha \Goes{P} \rho$ in the conservative semantics.
\end{thm}

\begin{proof}
By induction on the number of 1-$P$-unfoldings needed
to derive $\alpha'$ from $\alpha$.
The base case (0 unfoldings) is trivial.  For the inductive case
($n$ unfoldings), let $\alpha''$ be a 1-$P$-unfolding
of $\alpha$ such that $\alpha'$ is a $P$-unfolding of $\alpha''$
after $n-1$ unfoldings.  By the induction hypothesis,
$\theta: \alpha'' \Goes{P} \rho$ in the conservative semantics.

It remains to prove that $\theta: \alpha \Goes{P} \rho$ as well.
We do this by induction on the structure of the $\alpha''$
computation.  The cases are on the bottommost rule applied.
In all cases, if $\alpha$ starts with a predicate call and
$\alpha''$ is derived from it by unfolding that call, then
the computation of $\alpha$ can be derived from that of
$\alpha''$ by just adding an application of Pred.  Otherwise,
all cases follow directly from one or more applications of
the induction hypothesis.
\end{proof}

This property of predicate unfoldings and the pessimistic
semantics will be useful for the rest of the paper, because it
allows us to abstract away from the unbounded computations of
the non-pessimistic semantics and consider only the simpler,
bounded computations of the pessimistic semantics.

\subsubsection{The Definedness Ordering}

The following definitions and theorem makes the connections
between unfoldings and the pessimistic semantics more precise
and concise by allowing us to give an expression corresponding
to the outcome of a goal in terms of its pessimistic outcome.

\begin{figure}[tp]

\centerline{\epsffile{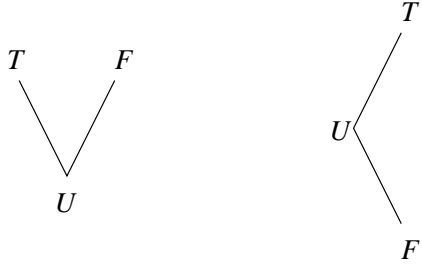}}

\caption{
  Hasse diagrams of the ``definedness'' ordering $<_k$ (left)
  and the ``truth'' ordering $<_t$ (right) of truth values.
}
\label{hasse-fig}
\end{figure}

We define the {\it definedness ordering} $<_k$ on truth values
as the least partial order relation such that $U <_k T$ and $U <_k F$
(see Figure \ref{hasse-fig}).  This is a standard ordering for
these three truth values; see for example \cite{belnap-four-valued}.
The expression $max_k(S)$, where $S$ is a set of truth values,
is undefined if $\{T, F\} \subseteq S$, and otherwise
is defined as the unique truth value $V$ such that $W \leq_k V$
for all $W \in S$.

Finally, we give the first raising lemma.

\begin{lemma}[Raising Lemma 1]
\label{unf-outcome-char-thm}
  For any goal $G$,
  $max_k(\{outcome^\frown(G') ~|~ G'$ is a
  $P$-unfolding of $G\})$
  is well-defined and equal to $outcome_P(G)$.
\end{lemma}

\begin{proof}
Let the set $S$ of truth values be
$\{outcome^\frown(G') ~|~ G'$ is a $P$-unfolding of $G\}$.
First assume that $outcome_P(G) = T$.
By Theorem \ref{cons-unf-thm}, $T \in S$;
however, if $F \in S$, then by Theorem \ref{unf-cons-thm},
$outcome_P(G) = F$, a contradiction.  Therefore
$max_k(S)$ is defined and must be $T$.
Similarly, if $outcome_P(G) = F$ then $max_k(S)$ is defined
and equal to $F$.
If $outcome_P(G) = U$, then it cannot be the case that
$T \in S$ or $F \in S$, because otherwise, by Theorem
\ref{unf-cons-thm}, $outcome_P(G) \not= U$.  Therefore
$S = \{U\}$, and $max_k(S)$ is defined and equal to $U$.  
\end{proof}

\subsection{Depth-First Normal Form}
\label{dfnf-section}

We now turn to the notion of depth-first normal form (DFNF)
in order to increase the level of abstraction of the semantics.  
The DFNF of a formula $G$ is a formula which
is operationally equivalent to $G$ but whose outcome can be
given a compositional characterization.
In this section, we first define a term-rewriting system which
rewrites formulas into formulas.  We then prove that the system
is locally confluent and terminating, and that it transforms
every formula to a unique normal form (which we define as the
DFNF).  We then prove that each of the transformations of the
rewriting system preserves pessimistic outcome.  The conclusion
is that each goal has a unique DFNF which has the same
pessimistic outcome as the original goal.

The DFNF by itself does not directly raise the abstraction
level of the semantics; however, it puts a goal in a form which
can be given an abstract characterization, as we will see in the
next section.  The conclusion of this section is therefore referred
to as the second raising lemma.

\subsubsection{Term-Rewriting System}

The notion of DFNF, which is closely related to the notion
of disjunctive normal form (DNF), was introduced in
\cite{andrews-lnaf-tcs}.  Here we expand the notion to take
account of $if$ formulas.

The classes of {\it negated-disjunction} ($N$) and
{\it outer-disjunction} ($O$) formulae
are defined mutually
recursively as follows.  (Informally, an $O$ formula has $\Or$s
directly inside only $\Not$s or other $\Or$s.)
\[ N ~~::=~~
    p(t_1, \ldots, t_n) ~~|~~
    s = t ~~|~~
    N \And N ~~|~~
    \exists x N ~~|~~
    \Not O
\]
\[ O ~~::=~~
    N ~~|~~
    O \Or O
\]
For example, $p \Or (\exists x(q(x)) \And r)$ is an
outer-disjunction formula but not a negated-disjunction formula;
however, $\Not (p \Or (\exists x(q(x)) \And r))$ is a
negated-disjunction formula and thus automatically an
outer-disjunction formula.

\begin{figure}[tp]

  \begin{itemize}
  \item[R1]
    $(B_1 \Or B_2) \And C ~\Rw{}~ (B_1 \And C) \Or (B_2 \And C)$
  \item[R2]
    $B \And (C_1 \Or C_2) ~\Rw{}~ (B \And C_1) \Or (B \And C_2)$,
    where $B$ is negated-disjunction
  \item[R3]
    $ \exists x (B_1 \Or B_2) ~\Rw{}~
     (\exists x B_1) \Or (\exists x B_2)$
  \item[R4]
    $if[\vec{x}]((B_1 \Or B_2), C) ~\Rw{}~
     if[\vec{x}](B_1, C)
     \Or
     (\Not(\exists \vec{x} B_1)
      \And if[\vec{x}](B_2, C))$
  \item[R5]
    $if[\vec{x}](B, C) ~\Rw{}~
     \exists \vec{x} (B \And C)$,
    where $B$ is negated-disjunction
  \end{itemize}

\caption{The rules of the term-rewriting relation $\Rw{}$.}
\label{rw-fig}
\end{figure}

The notion of depth-first normal form is based on the five rules
R1-R5 of
the term-rewriting relation $\Rw{}$ (Figure \ref{rw-fig}), which
can be applied anywhere in a formula to rewrite it into another
formula.  Two of the rules refer to the notion of a
negated-disjunction formula.  We define a formula to be
{\it in depth-first normal form} if none of R1-R5 can be applied
anywhere in the formula.

For example, the formula
$if[x](x=0, p(x) \Or q(x))$
can be rewritten by one application of R5 to
$\exists x (x=0 \And (p(x) \Or q(x)))$,
and then by one application of R2 to
$\exists x ((x=0 \And p(x)) \Or (x=0 \And q(x)))$.
It can then be rewritten by one application of R3 to
$\exists x (x=0 \And p(x)) \Or \exists x (x=0 \And q(x))$.
None of the rules R1-R5 apply to this latter formula,
so it is in depth-first normal form.
\label{dfnf-example-section}

\subsubsection{Local Confluence and Termination}

To prove that the rewriting process always leads to a
single formula, we prove local confluence and termination of this
rewriting system.  The proofs are contained in Appendix
\ref{proofs-appx}.

\begin{thm}[Local Confluence of Rewriting System]
\label{local-confluence-thm}
  If $A \Rw{} A_1$ and $A \Rw{} A_2$,
  then there is an $A_3$ such that $A_1 \Rw{}^* A_3$ and $A_2 \Rw{}^* A_3$.
\end{thm}

\begin{proof}
See Appendix \ref{proofs-appx}.  
\end{proof}

In preparation for proving termination of the rewriting system, we
define the {\it depth} $d(G)$ of a formula $G$.  It is the
conventional notion of depth of a formula, expanded to take
account of $if$. \\
\begin{tabular}{l}
  $d(s=t) = d(p(t_1, \ldots, t_n)) = 1$ \\
  $d(B \And C) = d(B \Or C) = max(d(B),d(C)) + 1$ \\
  $d(\Not B) = d(\exists x~B) = d(B) + 1$ \\
  $d(if[x_1,\ldots,x_n](B, C)) = max(d(B),d(C)) + 1$ \\
\end{tabular}

We also define the {\it maximum potential depth} $pd(G)$ of a
formula $G$.  This is the depth that the formula might possibly
attain after repeatedly being transformed with R1-R5.  \\
\begin{tabular}{l}
  $pd(s=t) = pd(p(t_1, \ldots, t_n)) = 1$ \\
  $pd(B \And C) = pd(B \Or C) = max(pd(B),pd(C)) + 1$ \\
  $pd(\Not B) = pd(\exists x~B) = pd(B) + 1$ \\
  $pd(if[x_1,\ldots,x_n](B, C)) = n + 2pd(B) + max(pd(B), pd(C))$ \\
\end{tabular} \\
Clearly $1 \leq d(G) \leq pd(G)$ for all formulas $G$.

The main lemma we need for termination is to prove that each
application of R1-R5 maintains or decreases potential depth.

\begin{lemma}
\label{rule-decreases-pd-lemma}
  If $G \Rw{} G'$, then $pd(G) \geq pd(G')$.
\end{lemma}

\begin{proof}
See Appendix \ref{proofs-appx}.  
\end{proof}

\begin{thm}[Termination of Rewriting System]
\label{termination-thm}
  For every $G$, there is an integer $j$ such that for
  every sequence of formulas $G = G_0, G_1, G_2, \ldots, G_k$
  such that $G_i \Rw{} G_{i+1}$ for all $1 \leq i < k$,
  we have that $k \leq j$.
\end{thm}

\begin{proof}
Each of the rules R1-R5 increase the number of connectives
in the formula, where $if$ is counted as one connective.  However,
the Lemma shows that the depth of the resultant formula is
bounded by $pd(G)$.  Since the formula tree has a bounded depth and
bounded branching factor, there is a limit $j$ to how many nodes
(connectives) it can contain.  The rewriting process must stop
at or before this limit.  
\end{proof}

\subsubsection{Unique Normal Form and DFNF}

Because of local confluence and termination, we are able to
state the following corollary, which shows that every goal has a
unique normal form under the rewriting rules R1-R5.

\begin{cor}[Unique Normal Form]
\label{unique-normal-form-cor}
  For every formula $G$ not in normal form, there is a unique
  formula $G''$ in normal form, such that for all $G'$ such that
  $G \Rw{} G'$, we have that $G' \Rw{}^* G''$.
\end{cor}

\begin{proof}
See Appendix \ref{proofs-appx}.  
\end{proof}

Because of this corollary, we are justified in
making the following definition.  The {\it depth-first normal
form} of a formula, $dfnf(G)$, is the unique formula
$G'$ such that $G \Rw{}^* G'$ and there is no $G''$ such that
$G' \Rw{} G''$.
(For instance, the depth-first normal form of the example
formula from Section \ref{dfnf-example-section},
$if[x](x=0, p(x) \Or q(x))$, is
$\exists x (x=0 \And p(x)) \Or \exists x (x=0 \And q(x))$.)
Clearly, despite the complexity of the proofs
of confluence and termination, we can obtain $dfnf(G)$ in a
straightforward fashion, by simply applying one of the rules
R1-R5 to any suitable redex (say, the outermost one) until
there are no more redexes.

We also note that $dfnf(G)$ is outer-disjunction, a fact which
will be important soon.

\begin{theorem}
\label{dfnf-outcome-char-thm}
  For all $G$, $dfnf(G)$ is outer-disjunction.
\end{theorem}

\begin{proof}
If $dfnf(G)$ were not outer-disjunction, it would have
some disjunction as an immediate subformula of a conjunction,
existential formula, or $if$ formula.  In all these cases,
one of rules R1-R5 would apply. 
\end{proof}

\subsubsection{Outcome Preservation}

We now show that the depth-first normal form formation does
not change the outcome of a goal under the pessimistic
semantics.

\begin{thm}[General Result Preservation of $dfnf$]
\label{general-result-pres-thm}
  If $\alpha'$ is $\alpha$
  with some formulas transformed by applications of rules R1-R5,
  then $\theta: \alpha \Goes{} \rho$ in the pessimistic semantics iff
  $\theta: \alpha' \Goes{} \rho$ in the pessimistic semantics.
\end{thm}

\begin{proof}
See Appendix \ref{proofs-appx}.  
\end{proof}

We can now give the second raising lemma, by showing the
specific result that we wanted to obtain.

\begin{lemma}[Raising Lemma 2]
\label{outcome-pres-cor}
  $outcome^\frown(G) = outcome^\frown(dfnf(G))$.
\end{lemma}

\begin{proof}
By Theorem \ref{general-result-pres-thm},
with respect to the pessimistic semantics,
$((): G \Goes{} \rho)$ iff $((): dfnf(G) \Goes{} \rho)$.
Therefore, with respect to the pessimistic semantics,
$G$ succeeds (fails, diverges) exactly when $dfnf(G)$
succeeds (fails, diverges).
\end{proof}

Note that we have come one step closer to an abstract
characterization of outcome, by reducing the problem of
characterizing outcome of a general goal with respect to a
general program to the problem of characterizing the outcome of
an outer-disjunction goal with respect to the pessimistic
semantics.
 
\subsection{The Valuation Function}
\label{valuation-section}

Finally we come to the definition of the valuation $v$, which
characterizes the outcomes of outer-disjunction goals (e.g.,
goals in DFNF) with respect to the pessimistic semantics.
This valuation is a compositional function from formulae to
truth values, like valuations in standard theories of truth
\cite{kripke,fitting-kripke}, and interprets the binary connectives
in a manner consistent with the left-to-right search algorithm
of Prolog.  $v$ is based on the similar valuation in
\cite{andrews-lnaf-tcs}.  The valuation in that paper
is on a domain of
four truth values, but we need only three truth values here
because we do not consider the $flounder$ outcome.

In this section, we first define the alethic ordering $<_t$ on
truth values, and then the valuation function $v$ which uses it.
Then we show that the valuation of a goal in outer-disjunction
form is the same as its pessimistic outcome.

\subsubsection{Alethic Ordering and Valuation Function}

We define the {\it alethic ordering} $<_t$ on truth values as
the least partial order relation such that $F <_t U$ and $U <_t T$.
(See Fig. \ref{hasse-fig}.  This is another standard ordering on
these truth values; see for instance \cite{belnap-four-valued}.)
The expression $max_t(S)$, where $S$ is a set of truth values,
is defined as the unique truth value $V$ such that $W \leq_t V$
for all $W \in S$.  The alethic ordering is used in the
valuation function to express the meaning of $\exists x~G$ in
terms of the meaning of the instances of $G$.

$v$, a {\it valuation function} mapping ground,
outer-disjunction (O) formulae to truth values in
$\{T, U, F\}$, is defined as follows.
\begin{recdef}
\item $v(t=t) = T$;
\item $v(s=t) = F$, where $s$ is not identical to $t$;
\item $v(p(t_1, \ldots, t_n)) = U$;
\item \parbox{10cm}{
      \( v(B \And C) = \left\{\begin{array}{ll}
                                v(C) & \mbox{if $v(B)=T$,} \\
                                v(B) & \mbox{otherwise;}
                                \end{array}
                         \right. \)
      \hfill{}
      }
\item \parbox{10cm}{
      \( v(B \Or  C) = \left\{\begin{array}{ll}
                                v(C) & \mbox{if $v(B)=F$,} \\
                                v(B) & \mbox{otherwise;}
                                \end{array}
                         \right. \)
      \hfill{}
      }
\item $v(\exists x~B) = max_{t}(\{v(B[x:=t]) ~|~ t \mbox{ ground}\})$;
\item \parbox{10cm}{
      \( v(\Not B) =  \left\{  \begin{array}{ll}
                                F & \mbox{if $v(B)=T$,} \\
                                U & \mbox{if $v(B)=U$,} \\
                                T & \mbox{if $v(B)=F$.}
                                \end{array}
                       \right. \)
      \hfill{}
      }
\end{recdef}
For instance, recall from Section \ref{true-false-section}
that $true$ is the formula $(0=0)$ and
$false$ is the formula $(0=1)$.  By the definition of $v$,
we have that $v(true)=v(0=0)=T$, and $v(false)=v(0=1)=F$, as
expected.  We also have that $v(\Not true)=F$,
$v(true \And false) = F$, and $v(false \Or true) = T$.
We have that $v(false \Or p(0))$ and $v(true \And p(0))$ are both
$U$, but $v(false \And p(0))=F$ and $v(true \Or p(0))=T$,
consistent with how the pessimistic semantics would execute
the formulas as queries.

In fact, while $v(0=0)=T$, we have that $v(s=0)=F$ for any term
$s$ other than $0$.  Therefore the set
$\{v(t=0) ~|~ t \mbox{ ground}\}$ is the set
$\{v(0=0)\} \cup 
 \{v(t=0) ~|~ t \mbox{ ground and } t \neq 0\}$,
i.e.\ $\{T\} \cup \{F\}$, or $\{T, F\}$.
As a consequence,
$v(\exists x (x=0)) =
 \{v(t=0) ~|~ t \mbox{ ground}\} = T$,
since $T$ is the maximally true truth value in the set $\{T, F\}$.

\subsubsection{Equivalence of Valuation and Pessimistic Outcome}

The valuation function $v$ characterizes precisely the behaviour of
outer-disjunction formulae with respect to the pessimistic
semantics.  In preparation for this result, we state a
proposition which is a weaker form of the converse of the
witness properties, applying only to $N$ formulas.

\begin{prop}
\label{reverse-witness-prop}
  Let $\alpha$ be a sequence of negated-disjunction ($N$)
  formulas, such that $\theta: \alpha \Goes{} \rho$ in the
  pessimistic semantics.
  Let $V$ be a subset of the free variables of $\alpha$.
  Then: \\
  (1) If for some substitution $\xi$ grounding $V$ consistent
    with $\theta$, $(\theta: \alpha\xi \Goes{} \theta')$ in the
    pessimistic semantics, then $\rho$ is some $\theta''$. \\
  (2) If for all substitutions $\xi$ grounding $V$ consistent
    with $\theta$, $(\theta: \alpha\xi \Goes{} fail)$ in the
    pessimistic semantics, then $\rho$ is $fail$.
\end{prop}
The fragment of the pessimistic semantics
dealing with negated-disjunction formulas is identical to
the fragment of the semantics of \cite{andrews-lnaf-tcs}
dealing with negated-disjunction formulas with respect to
the empty program.  The proof of this Proposition is thus
a simple adaptation of
the proof of Lemma 4.5 from \cite{andrews-lnaf-tcs}.
Intuitively, the Proposition applies only to $N$
formulas because instantiating an $N$ formula will
either cause it to fail or will not change the outcome
its computation.  In contrast, for example, $B \Or C$
may diverge because $B$ diverges, but $B\theta \Or C\theta$
may succeed because $B\theta$ fails
and $C\theta$ succeeds.  We cannot draw any conclusions about
the behaviour of $B \Or C$ from the behaviour of its instances.

We are now in a position to state the third raising lemma,
continuing our process of abstraction.  Note that this lemma
relates an operational notion (pessimistic outcome)
to an entirely abstract one (valuation).

\begin{theorem}[Raising Lemma 3]
\label{v-outcome-char-thm}
  If $G$ is ground and
  outer-disjunction, then $v(G) = outcome^\frown(G)$.
\end{theorem}

\begin{proof}
By induction on the structure of $G$.  Cases are on the
outermost connective.
We note only the three subcases of the
case in which $G = \exists x~B$.

If $outcome^\frown(G) = T$, there must be some
$\theta'$ such that $((): \exists x~B \Goes{} \theta')$ in
the pessimistic semantics.
In this case, we also have that
$((): B[x:=x'] \Goes{} \theta')$, and by the witness properties,
there must be some ground $t$ and $\theta''$ such that
$((): B[x:=x'][x':=t] \Goes{} \theta'')$.
Thus for some $t$, $outcome^\frown(B[x:=t]) = T$.
By the induction hypothesis,
$v(B[x:=t]) = T$; and by the definition of $max_t$, $v(G) = T$.

If $outcome^\frown(G) = F$, then
$((): \exists x~B \Goes{} fail)$ in
the pessimistic semantics.
In this case, we also have that
$((): B[x:=x'] \Goes{} fail)$, and by the witness properties,
for all ground $t$,
$((): B[x:=x'][x':=t] \Goes{} fail)$.
Thus for all $t$, $outcome^\frown(B[x:=t]) = F$.
By the induction hypothesis,
$v(B[x:=t]) = F$; and by the definition of $max_t$, $v(G) = F$.

Otherwise, $outcome^\frown(G)=U$.
By Prop.\ \ref{reverse-witness-prop}, there cannot be
any $t$ such that $outcome^\frown(B[x:=t]) = T$, because otherwise
$outcome^\frown(G)$ would be $T$; and again
by Prop.\ \ref{reverse-witness-prop}, it cannot be the case that
for all $t$, $outcome^\frown(B[x:=t]) = F$, because otherwise
$outcome^\frown(G)$ would be $F$.  Thus for some $t$,
$outcome^\frown(B[x:=t]) = U$, so the set
$\{v(B[x:=t]) ~|~ t$ is ground $\}$ of truth values is either
$\{U\}$ or $\{U, F\}$.
Thus by the definition of $max_t$, $v(G) = U$.
\end{proof}

\subsection{The Denotation of a Program}
\label{denotation-section}

In this section, we give the final raising lemma which summarizes
the previous ones.  This lemma gives an expression which is an
abstract characterization of the outcome of a goal; we therefore
give a definition of the denotation of a program which uses this
expression.

\begin{lemma}[Raising Lemma 4]
\label{v-top-outcome-char-thm}
  For any ground goal $G$, \\
  $outcome_P(G) = max_k(\{v(dfnf(G')) ~|~ G'$ is a
  $P$-unfolding of $G\})$.
\end{lemma}

\begin{proof}
By Raising Lemma 1,
$outcome_P(G) = max_k(\{outcome^\frown(G') ~|~ G'$ is a $P$-unfolding of $G\})$.
By Raising Lemma 2, $outcome^\frown(G') = outcome^\frown(dfnf(G'))$ for
any $G'$.  But by Theorem \ref{dfnf-outcome-char-thm},
$dfnf(G')$ is in outer-disjunction form for any $G'$; therefore by Raising
Lemma 3, $outcome^\frown(dfnf(G')) = v(dfnf(G'))$.
Putting this all together, we conclude that
$outcome_P(G) = max_k(\{v(dfnf(G')) ~|~ G'$ is a $P$-unfolding of $G\})$.
\end{proof}

We therefore make the following
definition.  The {\it denotation} $v_P$ of a program $P$ is a
valuation function defined by:
\begin{center}
  $v_P(G) = max_k(\{v(\dfnf(G')) ~|~ G'$ is an unfolding of $G\})$.
\end{center}

We have the following trivial theorem.
\begin{theorem}[Denotation]
  For any ground goal $G$, $outcome_P(G) = v_P(G)$.
\end{theorem}

\begin{proof}
By Raising Lemma 4 and the definition of $v_P$.
\end{proof}

Note that the restriction to ground goals does not decrease the
generality of the denotation result, since a goal $G$ with free
variables $\vec{x}$ has the same outcome as the goal
$\exists \vec{x} G$.

\subsection{Example}
\label{sems-example-section}

As a further example of how the denotation of a program defines the
correct truth value of a goal, we derive the value obtained by
applying the denotation of a program to a goal.

Let the program $P$ be the second
``delete'' program from Section \ref{example-section}:
\begin{tabbing}
\Indent{} \= $\Or$ \= \kill
\>  $d(x, y, z) \colondash$ \\
\>  \>  $(y=[~] ~ \And ~ z=[~])$ \\
\>  $\Or$ \> $if[ys](y=[x|ys], d(x, ys, z))$ \\
\>  $\Or$ \> $(\Not\exists ys(y=[x|ys]) ~ \And$ \\
\>        \> \Indent{} $\exists y' \exists ys \exists zs(y=[y'|ys] \And z=[y'|zs] \And d(x, ys, zs)))$
\end{tabbing}
Consider the goal $G = \exists z ~ d(a, [~], z)$.
This goal asks whether
there is a $z$ which is obtained by deleting $a$ everywhere from
the empty list $[~]$.  It has the outcome
$T$ in the conservative semantics, since there does exist a
$z$, namely the empty list $[~]$ itself, which is obtained that way.

We take as our objective to derive the value of $v_P(G)$.
From the definition of $v_P$, we have that
$v_P(G) = max_k(\{v(\dfnf(G')) ~|~ G'$ is an unfolding of $G\})$.
Let $S$ be the set 
$\{v(\dfnf(G')) ~|~ G'$ is an unfolding of $G\}$; then
$v_P(G) = max_k(S)$.  As discussed in the proof of Raising Lemma 1,
if $\{U, T\} \subseteq S$, then $F \not\in S$; so if we can find
one unfolding of $G$ whose DFNF valuation is $U$ and another
whose DFNF valuation is $T$, then we know $S = \{U, T\}$.

In fact, we can find such unfoldings.  The subsequent sections
show that $G$ itself is such that $v(dfnf(G)) = U$, and that
the first unfolding $G_1$ of $G$ is such that $v(dfnf(G_1)) = T$.
Hence $v_P(G) = max_k(S) = max_k(\{U, T\}) = T$.

First, we show that $v(dfnf(G)) = U$.  Then, we find the expression
for $G_1$ and for $dfnf(G_1)$.  Finally, we show that $v(dfnf(G_1)) = T$.

\subsubsection{$v(dfnf(G)) = U$}

$G$ is $\exists z ~ d(a, [~], z)$.  This formula contains no
disjunctions or $if$s, so none of the DFNF rewriting rules applies
to it; hence $dfnf(G)$ is $G$ itself.  By the definition of $v$,
$v(dfnf(G)) = v(G) = v(\exists z ~ d(a, [~], z))$, which is the expression
$max_t(\{ d(a, [~], z) ~|~ t$ is a ground term$\})$; that is,
the maximally true truth value amongst the valuations of all
the formulas of the form $d(a, [~], t)$, where $t$ is a ground term.

However, by the definition of $v$, the valuation of any
predicate call formula is $U$ (since $v$ correctly characterizes
the pessimistic semantics).  Hence 
$max_t(\{ d(a, [~], z) ~|~ t$ is a ground term$\}) = max_t(\{U\}) = U$.
Since this was the expression for $v(dfnf(G))$, we have that
$v(dfnf(G)) = U$.

\subsubsection{First Unfolding and its DFNF}

$G$ is $\exists z ~ d(a, [~], z)$.
The first unfolding of $G$, $G_1$, can be obtained by replacing
the predicate call within it by the body of the definition of
the predicate $d$, replacing formal by actual parameters.
Therefore:
\begin{tabbing}
\Indent{} \= $\Or$ \= \kill
\>  $G_1 = \exists z($ \\
\>  \>  $([~]=[~] ~ \And ~ z=[~])$ \\
\>  $\Or$ \> $if[ys]([~]=[a|ys], d(a, ys, z))$ \\
\>  $\Or$ \> $(\Not\exists ys([~]=[a|ys]) ~ \And$ \\
\>        \> \Indent{} $\exists y' \exists ys \exists zs([~]=[y'|ys] \And z=[y'|zs] \And d(a, ys, zs)))$
\end{tabbing}
We abbreviate this formula as
$\exists z(G'_1 \Or G'_2 \Or (G'_3 \And G'_4))$.

The DFNF rewriting rule R3 can be applied twice to $G_1$, to
yield the formula 
$(\exists z(G'_1) \Or \exists z(G'_2) \Or \exists z(G'_3 \And G'_4))$.
$G'_2$ is an $if$ formula,
$if[ys]([~]=[a|ys], d(a, ys, z))$, whose first subformula
$([~]=[a|ys])$ is a negated-disjunction formula; hence, the
DFNF rewriting rule R5 can be applied to it, yielding the
subformula $G'_5 = \exists ys([~]=[a|ys] \And d(a, ys, z))$.
At this point, no more of the DFNF rewriting rules can be
applied to the formula, so it is in depth-first normal form.

Hence, $dfnf(G_1) = 
(\exists z(G'_1) \Or \exists z(G'_5) \Or \exists z(G'_3 \And G'_4))$,
where:
\begin{itemize}
\item $G'_1 = ([~]=[~] ~ \And ~ z=[~])$;
\item $G'_5 = \exists ys([~]=[a|ys] \And d(a, ys, z))$;
\item $G'_3 = \Not\exists ys([~]=[a|ys])$; and
\item $G'_4 = \exists y' \exists ys
              \exists zs([~]=[y'|ys] \And z=[y'|zs] \And d(a, ys, zs)))$.
\end{itemize}

\subsubsection{$v(dfnf(G_1)) = T$}

$v(dfnf(G_1)) =
 v(\exists z(G'_1) \Or \exists z(G'_5) \Or \exists z(G'_3 \And G'_4))$.
We can therefore obtain the value of $v(dfnf(G_1))$ by first obtaining
the values of its disjuncts.
By the definition of $v$,
we have that $v(\exists z(G'_1))$ is the value of the expression
$max_t(\{v([~]=[~] ~ \And ~ t=[~]) ~|~ t$ is a ground term$\})$.
The value of $v([~]=[~] ~ \And ~ t=[~])$ is $T$ if the values of both
$v([~]=[~])$ and $v(t=[~])$ are $T$, and it is $F$ otherwise.
However, $v([~]=[~])$ is always $T$; and
$v(t=[~])$ is $T$ if $t$ is $[~]$, and otherwise is $F$.

The set
$\{v([~]=[~] ~ \And ~ t=[~]) ~|~ t$ is a ground term$\}$
therefore consists of the two truth values $\{T, F\}$.  The maximally
true member of this set is $T$; hence,
$v(\exists z(G'_1)) = max_t(\{T, F\}) = T$.
Now, $dfnf(G_1)$ is of the form $(\exists z(G'_1) \Or H)$;
so $v(dfnf(G_1)) = v(\exists z(G'_1) \Or H)$.  By the
definition of $v$, and because $v(\exists z(G'_1)) = T$,
$v(\exists z(G'_1) \Or H) = T$; hence $v(dfnf(G_1)) = T$.

We conclude the example by reiterating the value of $v_P(G)$.
Because $v(dfnf(G)) = U$ and $v(dfnf(G_1)) = T$, the set
$\{v(\dfnf(G')) ~|~ G'$ is an unfolding of $G\}$ is just
$\{U, T\}$.  Therefore: \vspace{2mm} \\
\begin{tabular}{lll}
$v_P(G)$ & $=$ & $max_k(\{v(\dfnf(G')) ~|~ G'$ is an unfolding of $G\})$ \\
         & $=$ & $max_k(\{U, T\})$ \\
         & $=$ & $T$ \\
\end{tabular} \\
This result accords with the fact that the original goal $G$
did succeed under the conservative semantics.

\subsection{Discussion}
\label{sems-discussion-section}

Note that the abstract semantics is based on six basic,
relatively simple notions: the notion of truth value,
the two orderings of the truth
values, the notion of predicate unfolding, the notion of
depth-first normal form, and the logical valuation.
The notion of depth-first normal form, in turn, is based on a
rewriting system of five rules.  The predicate unfolding and
normal form constructions essentially do local
meaning-preserving transformations to prepare the goal in
question for characterization, and the valuation actually
performs that characterization.

In some sense, the crucial element of the abstract semantics,
the element which allows it not to reify such notions as
substitutions and unification, is the $\exists$ clause of the
definition of $v$.  Rather than view a variable operationally,
as a placeholder in a term which at some future point can be
replaced by another term, the $\exists$ clause allows us to view
it as a true variable ranging over a fixed domain of discourse.
This, in turn, has been enabled by the
witness properties of the conservative semantics.  Without the
witness properties, we would not have been able to prove that
the value of $v(\exists x G)$ could be derived directly from the
consideration of the values of $v(G[x:=t])$, for any ground $t$.
Hence, the witness properties are useful not only from the point
of view of intuitively justifying the behaviour of a logic
programming system, but also on theoretical grounds.

\section{Conclusions}

The main contributions of this paper are as follows.
\begin{itemize}
\item We have defined an extension of Prolog with hard cut and
  negation as failure in which programs can provably be put
  in a convenient ``completed'' form.  This completion has
  been achieved by using a variable-binding choice construct, $if$.
\item We have identified the witness properties as important
  properties intermediate between the strict logicalness of pure
  Horn clause programming and the unrestricted freedom of
  typical Prolog implementations.
\item We have defined restrictions on the computation of
  extended programs which allow the resulting system to achieve
  the witness properties.  We have referred to the resulting
  notion of cut as {\it firm cut}, insofar as it is intermediate
  between hard and soft cut.
\item We have defined an abstract semantics for the restricted system
  (taking depth-first termination, rather than universal termination,
  as its observable),
  which uses the witness properties in order to avoid reifying
  the concepts of unification and substitution.
\end{itemize}

Long investigations by the author have not resulted in any
semantics for Prolog which allow the full range of behaviour of
hard cut while rising in any meaningful way above the level of
an operational semantics.  We do not believe at this point that
such a semantics is possible.  We believe that the system with
firm cut, as defined in this paper, is the best compromise
yet found between
the power of the hard cut and the logical rigour of the soft
cut.  We believe that the behaviours of hard cut excluded by
firm cut are unlikely to be missed by Prolog programmers, and
that the witness properties achieved by firm cut capture the
core of programmers' desiderata about a logic programming
system, even though they are not in complete harmony with logic.
However, these are merely beliefs.  We invite readers to decide
whether they agree or disagree based on their experience.

The more theoretically substantiated conclusions we draw from
this work are as follows.
\begin{itemize}
\item The widely-held view that features such as cut and
  negation as failure entirely destroy the declarative
  interpretation of logic programming systems seems to be too
  strong.  While firm cut cannot be interpreted as a logical
  construct, the abstract semantics developed here suggest that a
  system with firm cut is more declarative than one with hard cut,
  while still retaining behaviour of hard cut which is useful in
  practice.
\item If a logic programming language does not achieve soundness
  with respect to traditional logical interpretations, it might
  still be possible for it to achieve the witness properties.
  Given that practical, widely-used languages often implement
  pragmatic features which depart from well-defined semantics,
  insisting on the witness properties might be an acceptable
  alternative to insisting on soundness with respect to first
  order logic.
\item The Prolog syntax and clause-based operational semantics is
  difficult to work with in an abstract setting when taking cut into
  consideration.  We have found it easier to study semantic issues
  with programs in ``completed'' program form, and the
  structured operational semantics, described in this paper.  The
  syntax of the Mercury language \cite{mercury-jlp} is already
  closer to the completed form described here, since it uses an
  efficient ``if'' formula (though the ``if'' of Mercury
  corresponds to soft cut, not firm cut).
\end{itemize}

There are several interesting open questions suggested by
this research.
\begin{itemize}
\item Are other ``non-logical'' features of Prolog able to be
  given a form which allows the witness properties to be preserved?
  Obviously there is no hope for the {\tt var} and
  {\tt nonvar} predicates, which check the instantiation of their
  arguments, but what about {\tt assert}, {\tt retract},
  {\tt bagof}, and so on?
\item What is the largest subset of the liberal general
  semantics with the witness properties?  That is, can we define
  an operational semantics analogous to the conservative
  semantics, but with respect to which all goals with the witness
  properties do not
  flounder?  The answer to this question may lie with different
  strategies for coping with negation.
\item Can a mode {\it inference} system be devised which ensures
  non-floundering of goals?  That is, can we automate the process of
  defining modes for a program that will guarantee that no goal
  consistent with the inferred modes of the program's predicates
  will flounder?
\end{itemize}

We have implemented the ideas contained in this paper in an
experimental proof assistant program called SVP (Spreadsheet
Verifier for Prolog), whose user interface has been described
in \cite{andrews-spreadsheet-uitp}.  SVP transforms a Prolog
program with cuts into completed form, and then assists the user
in proving theorems in an assertion language similar to those
defined in \cite{andrews-phd-dd,staerk-lptp-jlp}.  We hope to
report on this work in the future.

\section{Acknowledgements}

Thanks to Ver\'{o}nica Dahl and the Logic and Functional
Programming Laboratory at Simon Fraser University for the use of
their facilities in preparing this material.  Thanks also to Kai
Salomaa for clarification on terminology.  Michel Billaud,
Robert \Staerk{}, and Torkel Franzen helped with the original
conference version of this paper, and the anonymous journal
referees contributed valuable comments and corrections.
This research is supported by
an NSERC (Natural Sciences and Engineering Research Council of
Canada) Individual Research Grant.

\bibliography{tplp}

\setcounter{section}{0}
\appendix
\newpage

\section{Proofs of Results}
\label{proofs-appx}

\subsection{Completion Algorithm Properties}

\noindent
{\bf Lemma \ref{completed-goal-stack-lemma}.}
\begin{quote} \it
  Let $\alpha$ be a goal stack.
  Let $\alpha'$ be $\alpha$
  with any number of occurrences of a sequence $B, C$ in a goal stack or
  clause body replaced by $B \And C$, where $B$ and $C$ are
  formulas.
  Then $(\theta: \alpha \Goes{P} \rho)$ in the liberal general semantics
  iff  $(\theta: \alpha' \Goes{P} \rho)$ in the liberal general semantics.
\end{quote}

\begin{proof}
By induction on the number of replacements of $B, C$ by
$B \And C$.  The base case (0 replacements) is trivial.  For the
inductive case, it suffices to demonstrate the case where
$\alpha'$ is derived from $\alpha$ by one replacement of $B, C$
by $B \And C$.  This in turn we prove by induction on the
structure of the computation of $\alpha$.
If $\alpha$ begins with $B, C$ and $\alpha'$ begins
with $B \And C$, then the computation of $\alpha'$ can be derived
from that of $\alpha$ with one Conj step.  Otherwise, either the
first formulas in the two goal stacks are identical, or they
have the same top-level connective; in either case, regardless
of the bottommost rule applied, the result
follows straightforwardly from the induction hypothesis.
\end{proof}

\noindent
{\bf
  Lemma \ref{conjoin-lemma}.
}
\begin{quote} \it
  Let $P'$ be $P$ with some sequence $B, C$ in a clause body
  replaced by $B \And C$.
  Then $\theta: \alpha \Goes{P} \rho$ in the liberal general semantics
  iff  $\theta: \alpha \Goes{P'} \rho$ in the liberal general semantics.
\end{quote}

\begin{proof}
By the Lemma, we can add new rules to the operational
semantics as follows:
\[ (1) \frac{\theta: \alpha' \Goes{P} \rho}
	{\theta: \alpha \Goes{P} \rho}
   \Sep
   (2) \frac{\theta: \alpha \Goes{P} \rho}
	{\theta: \alpha' \Goes{P} \rho}
\]
where $\alpha'$ is $\alpha$ with any number of occurrences of a
sequence $B, C$ in a goal stack or clause body replaced by
$B \And C$.  Moreover, by the
Lemma, we can essentially insert applications of these rules anywhere in a
computation and derive a computation of the premise from the
computation of the conclusion.

Therefore the ($\rightarrow$)
direction of the theorem can be proven as follows.  Given the
computation of $\theta: \alpha \Goes{P} \rho$, insert an
application of (1) above each Pred rule involving the clause
transformed in $P'$, obtaining the computation of the premise
from the Lemma.
The transformed proof will have sections of the form:
\begin{center}
\begin{tabular}{c}
  $\theta: p(\vec{t})using(\gamma'), \alpha \Goes{P} \rho$ \\
  $\overline{\underline{
    \theta: p(\vec{t})using(\gamma), \alpha \Goes{P} \rho
  }}$ \\
  $\theta: p(\vec{t}), \alpha \Goes{P} \rho$ \\
\end{tabular}
\end{center}
To obtain the computation of $\theta: \alpha \Goes{P'} \rho$,
replace each such section by 
\begin{center}
\begin{tabular}{c}
  $\underline{
    \theta: p(\vec{t})using(\gamma'), \alpha \Goes{P'} \rho
  }$ \\
  $\theta: p(\vec{t}), \alpha \Goes{P'} \rho$ \\
\end{tabular}
\end{center}
and replace $P$ by $P'$ everywhere else.
The other
direction of the theorem can be proven by inverting this operation. 
\end{proof}

\noindent
{\bf
  Theorem \ref{result-preservation-thm}.
}
\begin{quote} \it
  The completion algorithm preserves result according to
  the liberal general operational semantics.  That is,
  if $P'$ is the completion of $P$, then
  $\theta: \alpha \Goes{P} \rho$ in the liberal general semantics iff
  $\theta: \alpha \Goes{P'} \rho$ in the liberal general semantics.
\end{quote}

\begin{proof}
We prove the theorem by proving that each of the
transformations preserves result.  In what follows, we will
refer to the original program as $P$ and the program after
the single transformation in question as $P'$.

Step 2.2:
Clearly the two computations are equivalent up to a renaming
of some of the variables involved in the computations.

Step 2.3:
It suffices to show that any application of any of the four
$using$ rules with $P$ correspond to parts of computations with
$P'$.  Consider an application of the Using/nocut/succ rule with
$P$, where the formula being considered is an application of
predicate $p$.  The bottommost portion of the computation is:
\begin{center}
\begin{tabular}{c}
  $\underline{\theta\xi: \eta\xi, \alpha\xi \Goes{} \theta'}$ \\
  $\vdots$ \\
  $\overline{\theta: (s_1=t_1), \ldots, (s_k=t_k),
            \ldots, (s_n=x_n), \eta, \alpha \Goes{} \theta'}$ \\
  $\overline{\theta: p(s_1, \ldots, s_n) using
     (p(t_1, \ldots, t_k, \ldots, x_n) \colondash \eta),
                                      \gamma), \alpha \Goes{} \theta'}$ \\
\end{tabular}
\end{center}
where $\xi$ is the substitution resulting from the unifications.
With $P'$, the bottommost portion of the computation is the following:
\begin{center}
\begin{tabular}{c}
  $\underline{\theta\xi': \eta\xi', \alpha\xi' \Goes{} \theta'}$ \\
  $\vdots$ \\
  $\overline{\theta: (s_1=t_1), \ldots, (s_k=x_k),
           \ldots, (s_n=x_n), (x_k=t_k), \eta, \alpha \Goes{} \theta'}$ \\
  $\overline{\theta: p(s_1, \ldots, s_n) using
     (p(t_1, \ldots, x_k, \ldots, x_n) \colondash (x_k=t_k), \eta),
                                      \gamma), \alpha \Goes{} \theta'}$ \\
\end{tabular}
\end{center}
where $\xi'$ is the substitution resulting from the unifications.
However, by the properties of unification, we can rearrange
the equality formulas in the judgement second from the bottom to read:
$(s_1=t_1), \ldots, (x_k=s_k), (x_k=t_k),
\ldots, (s_n=x_n)$.  This sequence makes it
clear that the result substitution $\xi'$ is identical to $\xi$.
The cases of the other Using rules are proven similarly.

Step 3:
See Lemma \ref{conjoin-lemma} just before this theorem.

Step 4:  Let $\alpha$ be a goal stack, and let $\alpha'$ be
$\alpha$ with the formula $true$ inserted anywhere in a sequence
of goal stack elements or body elements.  Then
$(\theta: \alpha \Goes{P} \rho)$
iff $(\theta: \alpha' \Goes{P} \rho)$, by a simple structural
induction.  We can then follow the same line of reasoning as in
Lemma \ref{conjoin-lemma} to conclude that inserting $true$ anywhere
in a clause body preserves result.

Step 5:
When a clause with two consecutive cuts appears, instances of
the Body/cut/succ rule will arise in which $\eta_1$ is empty;
that is, a portion of some computations will be of the form
\begin{center}
\begin{tabular}{c}
 ${\overline{\theta: \epsilon \Goes{} \theta}  \Sep
  \theta: body(\eta_2), \alpha \Goes{} \rho}$ \\
 \cline{1-1}
 ${\theta: body(!, \eta_2), \alpha \Goes{} \rho}$ \\
\end{tabular}
\end{center}
where the Success rule has been used at the left-hand premise.
When the program is transformed to remove the second cut,
this portion of the computation will be replaced by the single
judgement $(\theta: body(\eta_2), \alpha \Goes{} \rho)$.

Step 6:  See Step 4 above.

Step 7:  See Step 4 above.

Step 8:
The original computation may have applications of
the Body/cut/succ rules of the following form:
\begin{center}
\begin{tabular}{c}
 ${\theta: \eta_1 \Goes{} \theta'  \Sep
  \theta': body(\eta_2)\theta', \alpha\theta' \Goes{} \rho}$ \\
 \cline{1-1}
 ${\theta: body(\eta_1, !, \eta_2), \alpha \Goes{} \rho}$ \\
\end{tabular}
\end{center}
This part of the computation is replaced in the new computation
by the following sequence:
\begin{center}
\begin{tabular}{ccc}
$\theta[\vec{y}:=\vec{y}']: \eta_1 \Goes{} \theta'$ \\
\cline{1-1}
$\vdots$ \\ 
\cline{1-1}
$\theta: \vec{y}=\vec{y}', \eta_1[\vec{y}:=\vec{y}'] \Goes{} \theta'$ & ~~~~~ &
  $\theta': body(\eta_2)\theta', \alpha\theta' \Goes{} \rho$ \\
\cline{1-3}
\multicolumn{3}{c}{
  $\theta: q(\vec{y}) using (q(\vec{y}') \colondash \eta_1[\vec{y}:=\vec{y}'], !, \eta_2[\vec{y}:=\vec{y}']), \alpha \Goes{} \rho$
} \\
  \cline{1-3}
  \multicolumn{3}{c}{
  $\theta: q(\vec{y}), \alpha \Goes{} \rho$} \\
  \multicolumn{3}{c}{
  $\overline{\theta: body(q(\vec{y})), \alpha \Goes{} \rho}$} \\
\end{tabular}
\end{center}
Note that the substitution $[\vec{y}:=\vec{y}']$ has the effect
of restoring $\eta_1, \eta_2$ to their original naming.
We do not show $[\vec{y}:=\vec{y}']$ elsewhere since the
computations are equivalent up to renaming.

The original computation may also have applications of
Body/cut/fail, which are transformed similarly.

Step 9.2:
In computations with $P$, variables in the clause are renamed
apart at the appropriate applications of the Pred rule.  In
computations with $P'$, the $\vec{y}$ variables are bound and
therefore not renamed apart.  However, they become renamed apart
in Exists rule applications above the application of the Using
or Body rule in which they become part of the goal stack.

Step 9.3:
The original computation may have portions ending with applications
of the Using/cut/succ rule, of the form \\
\begin{center}
\begin{tabular}{ccc}
$\theta\xi: F\xi \Goes{} \theta'$ \\
\cline{1-1}
$\vdots$ & \Sep & $\theta': G\theta', \alpha\theta' \Goes{} \rho$ \\
\cline{1-1} \cline{3-3}
$\theta: \vec{t}=\vec{x}, F \Goes{} \theta'$
 & & $\theta': body(G)\theta', \alpha\theta' \Goes{} \rho$ \\
\cline{1-3}
\multicolumn{3}{c}{
$\theta: p(\vec{t}) using (p(\vec{x}) \colondash F, !, G), \alpha \Goes{} \rho$
} \\
\end{tabular}
\end{center}
where $\xi$ is the substitution
resulting from the unification of $\vec{t}$ with $\vec{x}$.
(Without loss of generality, to avoid confusion, we assume that
the free variables of the clause are different from those in
$\vec{t}$ and $\alpha$, and do not require renaming apart.)
The computation with respect to $P'$ will have this portion of
the computation replaced by the following:
\begin{center}
\begin{tabular}{ccc}
$\theta\xi: F\xi \Goes{} \theta'$ & \Sep &
$\theta': G\xi\theta', \alpha\xi \Goes{} \rho$ \\
\cline{1-3}
\multicolumn{3}{c}{
  $\underline{\theta\xi: if[\vec{y}](F, G)\xi, \alpha\xi \Goes{} \rho}$
} \\
\multicolumn{3}{c}{$\vdots$ } \\
\multicolumn{3}{c}{
  $\overline{\theta: \vec{t}=\vec{x}, if[\vec{y}](F, G), \alpha \Goes{} \rho}$
} \\
\cline{1-3}
\multicolumn{3}{c}{
  $\theta: p(\vec{t}) using (p(\vec{x}) \colondash if[\vec{y}](F, G)), \alpha \Goes{} \rho$
} \\
\end{tabular}
\end{center}
However, because the $\vec{x}$ are distinct and different from the
variables in $\alpha$, $\alpha\xi$ is just $\alpha$; and because
$\theta'$ has arisen from $\theta\xi$, $\xi\theta' = \theta'$.  Thus the
two judgements at the top of this portion of this computation
are the same as the two at the top of the portion of the
computation with respect to $P$.

The original computation may also have applications of Using/cut/fail,
which are transformed similarly.

Step 10.2:
The original computation may have portions ending in
applications of the Using/nocut/succ
rule, of the form
\begin{center}
\begin{tabular}{c}
$\underline{\theta\xi: G\xi, \alpha\xi \Goes{} \rho}$ \\
$\vdots$ \\
$\overline{\theta: \vec{t}=\vec{x}, G, \alpha \Goes{} \rho}$ \\
$\overline{\theta: p(\vec{t}) using (p(\vec{x}) \colondash G; p(\vec{x}) \colondash H), \alpha \Goes{} \rho}$ \\
\end{tabular}
\end{center}
where $\xi$ is the substitution resulting from the unification of
$\vec{t}$ with $\vec{x}$.  (Again, without loss of generality we
assume the free variables of the clauses are different from those
of the conclusion.)
The computation with respect to $P'$ will have this portion of
the computation replaced by the following:
\begin{center}
\begin{tabular}{c}
$\underline{\theta\xi: G\xi, \alpha\xi \Goes{} \rho}$ \\
$\vdots$ \\
$\overline{\theta\xi: \exists\vec{y}(G)\xi, \alpha\xi \Goes{} \rho}$ \\
$\underline{\overline{\theta\xi: \exists\vec{y}(G)\xi \Or H\xi, \alpha\xi \Goes{} \rho}}$ \\
$\vdots$ \\
$\overline{\theta: \vec{t}=\vec{x}, \exists\vec{y}(G) \Or H, \alpha \Goes{} \rho}$ \\
$\overline{\theta: p(\vec{t}) using (p(\vec{x}) \colondash \exists\vec{y}(G) \Or H), \alpha \Goes{} \rho}$ \\
\end{tabular}
\end{center}
The topmost judgements of these portions of the
proof are the same.

The original computation may also have applications of Using/nocut/fail,
which are transformed similarly.

Step 10.3:
The original computation may have portions ending in
applications of the Using/cut/succ rule, of the form
\begin{center}
\begin{tabular}{ccc}
$\theta\xi: F\xi \Goes{} \theta'$ \\
\cline{1-1}
$\vdots$ & \Sep &
$\theta': G\theta', \alpha\theta' \Goes{} \theta'$ \\
\cline{1-1} \cline{3-3}
$\theta: \vec{t}=\vec{x}, F \Goes{} \theta'$ & \Sep &
$\theta': body(G)\theta', \alpha\theta' \Goes{} \theta'$ \\
\cline{1-3}
\multicolumn{3}{c}{
  $\theta: p(\vec{t}) using (p(\vec{x}) \colondash F, !, G; p(\vec{x}) \colondash H), \alpha \Goes{} \rho$
} \\
\end{tabular}
\end{center}
where $\xi$ is the substitution resulting
from the unification of $\vec{t}$ and $\vec{x}$.
(Throughout, we assume the variables of the clauses are distinct
from the other variables in the computation.)
The computation
with $P'$ will have this portion replaced by the following:
\[\begin{array}{c}
\underline{\theta\xi: F\xi \Goes{} \theta' \Sep 
 \theta': G\xi\theta', \alpha\xi\theta' \Goes{} \rho} \\
{\theta\xi: if[\vec{y}](F,G)\xi, \alpha\xi \Goes{} \rho} \\
\overline{\underline{\theta\xi: if[\vec{y}](F,G)\xi \Or ((\Not\exists\vec{y}(F) \And H)\xi, \alpha\xi \Goes{} \rho}} \\
\vdots \\
\overline{\theta: \vec{t}=\vec{x}, if[\vec{y}](F,G) \Or ((\Not\exists\vec{y}(F) \And H), \alpha \Goes{} \rho} \\
\overline{\theta: p(\vec{t}) using (p(\vec{x}) \colondash if[\vec{y}](F,G) \Or ((\Not\exists\vec{y}(F) \And H)), \alpha \Goes{} \rho} \\
\end{array}\]
As in Step 9.3, because of the way the substitutions were formed, the topmost
judgements in this portion of the $P'$ computation are the same as
those at the top of the portion of the $P$ computation.

The original computation may also have portions ending in
applications of the Using/cut/fail rule, of the form
\begin{center}
\begin{tabular}{ccc}
& & $\underline{\theta\xi: H\xi, \alpha\xi \Goes{} \rho}$ \\
$\theta\xi: F\xi \Goes{} fail$ & \Sep &
$\vdots$ \\
\cline{1-1}
$\vdots$ & \Sep &
$\overline{\theta: \vec{t}=\vec{x}, H, \alpha \Goes{} \rho}$ \\
\cline{1-1} \cline{3-3}
$\theta: \vec{t}=\vec{x}, F \Goes{} fail$ & \Sep &
$\theta: p(\vec{t}) using (p(\vec{x}) \colondash H), \alpha \Goes{} \rho$ \\
\cline{1-3}
\multicolumn{3}{c}{
  $\theta: p(\vec{t}) using (p(\vec{x}) \colondash F, !, G; p(\vec{x}) \colondash H), \alpha \Goes{} \rho$
} \\
\end{tabular}
\end{center}
where $\xi$ is the substitution renaming the variables of the
first clause apart, and $\xi'$ is the substitution resulting
from the unification of $\vec{t}$ and $\vec{x}$.  The computation
with $P'$ will have this portion replaced by the following:
\begin{center}
\begin{tabular}{ccccc}
& ~ & 
  $\theta\xi: F\xi \Goes{} fail$ \\
\cline{3-3}
& ~ & 
  $\vdots$ \\
\cline{3-3}
& ~ & 
  $\theta\xi: \exists\vec{y}(F)\xi \Goes{} fail$
  & ~ & 
  $\theta\xi: H\xi, \alpha\xi \Goes{} \rho$ \\
\cline{3-5}
$\theta\xi: F\xi \Goes{} fail$ & ~ & 
\multicolumn{3}{c}{
  $\theta\xi: \Not\exists\vec{y}(F)\xi, H\xi, \alpha\xi \Goes{} \rho$
} \\
\cline{1-1}
$\theta\xi: if[\vec{y}](F,G)\xi, \alpha\xi \Goes{} fail$ & ~ & 
\multicolumn{3}{c}{
  $\overline{\theta\xi: (\Not\exists\vec{y}(F) \And H)\xi, \alpha\xi \Goes{} \rho}$
} \\
\cline{1-5}
\multicolumn{5}{c}{
  $\underline{\theta\xi: if[\vec{y}](F,G)\xi \Or (\Not\exists\vec{y}(F) \And H)\xi, \alpha\xi \Goes{} \rho}$
} \\
\multicolumn{5}{c}{
  $\vdots$
} \\
\multicolumn{5}{c}{
  $\overline{\theta: \vec{t}=\vec{x}, if[\vec{y}](F,G) \Or (\Not\exists\vec{y}(F) \And H), \alpha \Goes{} \rho}$
} \\
\multicolumn{5}{c}{
  $\overline{\theta: p(\vec{t}) using (p(\vec{x}) \colondash if[\vec{y}](F,G) \Or (\Not\exists\vec{y}(F) \And H)), \alpha \Goes{} \rho}$
} \\
\end{tabular}
\end{center}
The three judgements at the top of this $P'$ computation portion
consist of two instances of one of the judgements at the top of the
$P$ portion, and one instance of the other one.

Since all the individual transformations preserve result, we
conclude that the entire transformation process preserves result.
\end{proof}

\subsection{Witness Properties of Conservative Semantics}

\noindent
{\bf
  Lemma \ref{general-failure-property-lemma}.
}
\begin{quote} \it
  Let $\theta, \alpha$ be such that
  $(\theta: \alpha \Goes{} fail)$ in the conservative semantics.
  Then for any $\xi$,
  $(\theta: \alpha\xi \Goes{} fail)$ in the conservative semantics.
\end{quote}

\begin{proof}
By induction on the structure of the computation of
$(\theta: \alpha \Goes{} fail)$.
Cases are on the bottommost rule applied.

Unif/succ:  Let $\sigma$ be the mgu found in the rule.
If $\xi \subseteq \sigma$, then $s\xi$ and $t\xi$
are identical,
and the result follows from
the induction hypothesis (IH).
Otherwise, if $s\xi$ and $t\xi$ have mgu $\sigma'$, then
since $\sigma$ is an mgu of $s$ and $t$,
there must be some $\xi'$ such that $\xi\sigma' = \sigma\xi'$.
The result then follows from the IH.
Otherwise, $s\xi$ and $t\xi$ do
not unify, and the computation fails with a single Unif/fail step.

Unif/fail:  If $s\xi$ and $t\xi$ had a unifier $\sigma$, then $s$
and $t$ would have a unifier $\xi\sigma$.  Since $s$ and $t$ have
no unifier, the computation of $\theta: \alpha\xi \Goes{} fail$
also consists of just one Unif/fail step.

Success:  Cannot occur.

Conj, Disj/nofail, Disj/fail:  Directly from the IH.

Exists:  We have not required that the substitution $\xi$
substitutes a term for $x'$.  Therefore the result follows from
the IH.

Not/succ:  $B$ has no free variables, so the computation
$\theta: B\xi \Goes{} fail$ is the same as that for
$\theta: B \Goes{} fail$.

Not/fail:  Again, $B$ has no free variables, so the computation
of the left-hand premise is the same. The result then follows from the IH.

Not/flounder, Not/sub:  cannot occur.

If/succ: We must prove that
$\theta: if[\vec{x}](B, C\xi), \alpha\xi \Goes{} fail$.
($B$ has no free variables other than $\vec{x}$, and
$if$ binds the variables $\vec{x}$.  We assume without
loss of generality that $dom(\xi) \cap \{\vec{x}\} = \emptyset$.)
For this, it suffices to prove that, for some $\theta'$,
$\theta: B[\vec{x}:=\vec{x}'] \Goes{} \theta'$
(which it does by assumption), and that
$\theta': C\xi[\vec{x}:=\vec{x}']\theta', \alpha\xi \Goes{} fail$.
Because $\vec{x}'$ do not appear in the conclusion,
$C\xi[\vec{x}:=\vec{x}']\theta'$ is the same thing as
$C[\vec{x}:=\vec{x}']\theta'\xi$.
The result therefore follows from the induction hypothesis.

If/fail: $B$ has no free variables other than
the $\vec{x}$ variables, so the computation
$\theta: B[\vec{x}:=\vec{x}']\xi \Goes{} fail$ is the same as that for
$\theta: B[\vec{x}:=\vec{x}'] \Goes{} fail$.
By the hypothesis, this computation fails.

If/flounder, If/sub:  Cannot occur.

Pred:  Directly from the IH. 
\end{proof}

\noindent
{\bf
  Lemma \ref{substitution-monotonicity-lemma}.
}
\begin{quote} \it
  Let $\theta, \alpha$ be such that $\alpha\theta \equiv \alpha$ and
  $\theta: \alpha \Goes{} \theta'$ in the conservative semantics.
  Then $\theta' \subseteq \theta$.
\end{quote}

\begin{proof}
By induction on the structure of the computation.  The
only rule which modifies the substitution in the judgements is
the Unif/succ rule, which obviously produces a more specific
substitution.  All other cases are straightforward consequences
of the induction hypothesis.  
\end{proof}

\noindent
{\bf
  Lemma \ref{general-success-property-lemma}.
}
\begin{quote} \it
  Let $\theta, \alpha$ be such that
  $\theta: \alpha \Goes{} \theta'$ in the conservative semantics.
  Let $V$ be a subset of the free variables of $\alpha$.
  Then for any $\xi$ grounding $V$ consistent with $\theta'$,
  $\theta: \alpha\xi \Goes{} \theta'\xi$ in the conservative semantics.
\end{quote}

\begin{proof}
By induction on the structure of the computation.
Cases are on the bottommost rule.

Unif/success:  Let $\sigma$ be the mgu found in the
rule.  By substitution monotonicity, any $\xi$ grounding
$V$ consistent with $\theta'$ must also ground $V$ consistent
with $\sigma$.  Thus $\alpha\xi\sigma$ is the same as $\alpha\sigma\xi$,
and the result follows from the induction hypothesis (IH).

Unif/fail:  Cannot occur.

Success:  Trivial.

Conj, Disj/nofail:  Directly from the IH.

Disj/fail:
From the General Failure Property, we have that
$\theta: B\xi, \alpha\xi \Goes{} fail$.
From the IH, we have that
$\theta: C\xi, \alpha\xi \Goes{} \theta'\xi$.
The result follows in one Disj/fail step.

Exists:  Because $V$ is also a subset of the free variables of
$B[x:=x']$, the result follows from the IH.

Not/succ:  Cannot occur.

Not/fail:  Because $B$ has no free variables, $B\xi$ is the same as
$B$.  The result follows from the original left-hand premise and
from the IH.

Not/flounder, Not/sub:  Cannot occur.

If/succ:  We assume without loss of generality that
$dom(\xi) \cap \{\vec{x}\} = \emptyset$.  (We can do this because
the $\vec{x}$ variables are renamed and thus can be prevented from
appearing in $\theta'$.)  We must therefore prove that
$(\theta: if[\vec{x}](B, C\xi), \alpha\xi \Goes{} \theta'\xi)$.
By assumption, $\theta: B[\vec{x}:=\vec{x}'] \Goes{} \theta''$
for some $\theta''$.  By the IH,
$\theta'': C[\vec{x}:=\vec{x}']\theta''\xi, \alpha\xi \Goes{} \theta'\xi$.
By substitution monotonicity, $\xi$ must ground $V$ consistent
with $\theta''$ as well.  Thus
$C[\vec{x}:=\vec{x}']\theta''\xi$ is the same as
$(C\xi)[\vec{x}:=\vec{x}']\theta''$, and the result
follows in one If/succ step.

If/fail, If/flounder, If/sub:  Cannot occur.

Pred:  Directly from the IH.
\end{proof}

\subsection{Depth-First Normal Form Results}

\noindent
{\bf
  Theorem \ref{local-confluence-thm}.
}
\begin{quote} \it
  If $A \Rw{} A_1$ and $A \Rw{} A_2$,
  then there is an $A_3$ such that $A_1 \Rw{}^* A_3$ and $A_2 \Rw{}^* A_3$.
\end{quote}

\begin{proof}

There are five cases, one for each of the rules R1-R5 applied
to derive $A_1$ from $A$.
We will give only the argument for R1, since the arguments for
the rest are similar or simpler.
We write $A[B_1, \ldots, B_n]$ for a formula A with
distinguished subformulas $B_1, \ldots, B_n$, and
$A[C_1, \ldots, C_n]$ for that formula with the distinguished
$B_1, \ldots, B_n$ replaced by $C_1, \ldots, C_n$.

Let $A$ be $A[(B_1 \Or B_2) \And C]$, and $A_1$ be
$A[(B_1 \And C) \Or (B_2 \And C)]$.  If $A_2$ is derived from
applying R1 to the same location, the result is trivially
true.  $A_2$ cannot be derived from applying R2 to the same
location, because $(B_1 \Or B_2)$ is not negated-disjunction.
$A_2$ also cannot be derived from applying R3-R5 to the same
location.  We therefore have four subcases.
In the first three subcases,
$A_2$ may be one of
$A[(B'_1 \Or B_2) \And C]$,
$A[(B_1 \Or B'_2) \And C]$, or
$A[(B_1 \Or B_2) \And C']$.
In the first subcase, one step from either $A_1$ or $A_2$ will
lead to
$A[(B'_1 \And C) \Or (B_2 \And C)]$. The second subcase is similar.
In the third subcase, two steps from $A_1$ and one from $A_2$ will lead to
$A[(B_1 \And C') \Or (B_2 \And C')]$.
The final subcase is when $A$ can be written as
$A[(B_1 \Or B_2) \And C, D]$, $A_1$ is
$A[(B_1 \And C) \Or (B_2 \And C), D]$, and $A_2$ is
$A[(B_1 \Or B_2) \And C, D']$.  In this case, one step from
either $A_1$ or $A_2$ will lead to 
$A[(B_1 \And C) \Or (B_2 \And C), D']$. 
\end{proof}

\noindent
{\bf
  Lemma \ref{rule-decreases-pd-lemma}.
}
\begin{quote} \it
  If $G \Rw{} G'$, then $pd(G) \geq pd(G')$.
\end{quote}

\begin{proof}
Clearly rules R1-R3 maintain potential depth;
the difficult cases are R4 and R5.

Case R4:
If R4 was applied at the top level, then we have
$G  = if[\vec{x}]((B_1 \Or B_2), C)$ and
$G' = if[\vec{x}](B_1, C)
     \Or
     (\Not(\exists \vec{x} B_1)
      \And if[\vec{x}](B_2, C))$.
Let the length of $\vec{x}$ be $n$.
Now we have that
\[\begin{array}{llll}
  pd(G') & = & \multicolumn{2}{l}{pd(if[\vec{x}](B_1, C) \Or
                  (\Not(\exists \vec{x} B_1) \And if[\vec{x}](B_2, C))} \\

         & = & max( & pd(if[\vec{x}](B_1, C)+1, \\
         & & &        pd(\Not(\exists \vec{x} B_1) \And if[\vec{x}](B_2, C))+1) \\

         & = & max( & 1+n+2pd(B_1)+max(pd(B_1), pd(C)), \\
         & & &        pd(\Not(\exists \vec{x} B_1) \And if[\vec{x}](B_2, C))+1) \\

         & = & max( & 1+n+3pd(B_1),  1+n+2pd(B_1)+pd(C), \\
         & & &        pd(\Not(\exists \vec{x} B_1)+2, \\
         & & &        pd(if[\vec{x}](B_2, C))+2) \\

         & = & max( & 1+n+3pd(B_1),  1+n+2pd(B_1)+pd(C), \\
         & & &        3+n+pd(B_1), \\
         & & &        2+n+2pd(B_2)+max(pd(B_2), pd(C)) \\

         & = & max( & 1+n+3pd(B_1),  1+n+2pd(B_1)+pd(C), \\
         & & &        3+n+pd(B_1), \\
         & & &        2+n+3pd(B_2),  2+n+2pd(B_2)+pd(C)) \\

         & = & max( & 1+n+3pd(B_1),  1+n+2pd(B_1)+pd(C), \\
         & & &        2+n+3pd(B_2),  2+n+2pd(B_2)+pd(C)) \\
\end{array}\]
There are now two subcases.  Subcase 1: if $pd(B_1) > pd(B_2)$, then
\[\begin{array}{lll}
  pd(G) & = & pd(if[\vec{x}]((B_1 \Or B_2), C) \\
        & = & n + 2pd(B_1 \Or B_2) + max(pd(B_1 \Or B_2), pd(C)) \\
        & = & n + 2 + 2pd(B_1) + max(1+pd(B_1), pd(C)) \\
        & = & max(3+n+3pd(B_1), 2+n+2pd(B_1)+pd(C)) \\
\end{array}\]
and $pd(G')$ simplifies to $max(1+n+3pd(B_1),  1+n+2pd(B_1)+pd(C))$.
Thus if $pd(C) > pd(B_1)$, we have
\[ pd(G) =~ (2+n+2pd(B_1)+pd(C)) ~>~ (1+n+2pd(B_1)+pd(C)) ~= pd(G')
\]
and otherwise ($pd(C) \leq pd(B_1)$) we have
\[ pd(G) =~ (3+n+3pd(B_1)) ~>~ (1+n+3pd(B_1)) ~= pd(G')
\]
so in both cases, $pd(G) > pd(G')$.

Subcase 2: otherwise, $pd(B_2) \geq pd(B_1)$.  We have:
\[\begin{array}{lll}
  pd(G) & = & max(3+n+3pd(B_2), 2+n+2pd(B_2)+pd(C)) \\
\end{array}\]
and $pd(G')$ simplifies to $max(2+n+3pd(B_2), 2+n+2pd(B_2)+pd(C))$.
Thus if $pd(C) > pd(B_2)$, we have
\[ pd(G) =~ (2+n+2pd(B_2)+pd(C)) ~=~ (2+n+2pd(B_2)+pd(C)) ~= pd(G')
\]
and otherwise ($pd(C) \leq pd(B_1)$) we have
\[ pd(G) =~ (3+n+3pd(B_2)) ~>~ (2+n+3pd(B_2)) ~= pd(G')
\]
so in both cases, $pd(G) \geq pd(G')$.

Similarly, if R4 was applied not at the top level,
$pd(G) \geq pd(G')$, since if any subformula is transformed
to have lower potential depth, the whole formula has lower
potential depth.

If R5 was applied at the top level, we have
$pd(G) = pd(if[\vec{x}](B, C)) = n + 2pd(B) + max(pd(B), pd(C)) =
max(n+3pd(B), n+2pd(B)+pd(C))$, and
$pd(G') = pd(\exists \vec{x} (B \And C)) = 1+n+max(pd(B), pd(C)) =
 max(1+n+pd(B), 1+n+pd(C))$.
If $pd(B) > pd(C)$, then
\[
  pd(G) =~ n+3pd(B) ~>~ 1+n+pd(B) ~= pd(G')
\]
and otherwise
\[
  pd(G) =~ n+2pd(B)+pd(C) ~>~ 1+n+pd(C) ~= pd(G')
\]
Thus in both cases $pd(G) > pd(G')$.

Similarly, if R5 was applied not at the top level,
$pd(G) > pd(G')$.
\end{proof}

\noindent
{\bf
  Corollary \ref{unique-normal-form-cor}.
}
\begin{quote} \it
  For every formula $G$ not in normal form, there is a unique
  formula $G''$ in normal form, such that for all $G'$ such that
  $G \Rw{} G'$, we have that $G' \Rw{}^* G''$.
\end{quote}

\begin{proof}
Let $k$ be the length of the longest chain of
rewritings that starts with $G$ (by Theorem
\ref{termination-thm} we know that this bound exists).  We prove
the corollary by induction on $k$.  In the base case ($k=1$),
we know from Theorem \ref{local-confluence-thm} that there
can be at most one unique $G'$ such that $G \Rw{} G'$; hence,
$G''$ is this $G'$.  In the inductive case, if there is a
unique $G'$ such that $G \Rw{} G'$, the result follows from the
induction hypothesis.  If there is more than one, then for each
pair $G_1$ and $G_2$ such that $G \Rw{} G_1$ and $G \Rw{} G_2$,
by Theorem \ref{local-confluence-thm}
there is some $G_3$ such that $G_1 \Rw{}^* G_3$ and $G_2 \Rw{}^* G_3$.
But by the induction hypothesis, there is some unique normal
form not only of $G_3$ but also of $G_1$ and $G_2$.  Because
$G_1 \Rw{}^* G_3$, the normal form of $G_3$ must be the same as
that of $G_1$, and similarly for $G_2$.  Hence all $G'$
such that $G \Rw{} G'$ must have some unique normal form $G''$.
This therefore is the unique normal form of $G$.  
\end{proof}

\noindent
{\bf
  Theorem \ref{general-result-pres-thm}.
}
\begin{quote} \it
  If $\alpha'$ is $\alpha$
  with some formulas transformed by applications of rules R1-R5,
  then $\theta: \alpha \Goes{} \rho$ in the pessimistic semantics iff
  $\theta: \alpha' \Goes{} \rho$ in the pessimistic semantics.
\end{quote}

\begin{proof}
By induction on the structure of the assumption computation.
If the application of the rules has not changed
the top-level connective of the first formula in $\alpha$, then the result
follows by the induction hypothesis.  Otherwise, we have cases
according to which of R1-R5 was used to transform the top-level
connective of the first formula.

Cases R1-R3 are very similar to the proof in \cite{andrews-lnaf-tcs}
and will not be repeated here.

Case R4:  The two computations are
$(\theta: if[\vec{x}((B_1 \Or B_2), C), \alpha \Goes{} \rho)$ and
$(\theta: if[\vec{x}](B_1, C)
     \Or
     (\Not(\exists \vec{x} B_1)
      \And if[\vec{x}](B_2, C)), \alpha \Goes{} \rho)$;
we must show that each implies the other.
There are several subcases.

If $(\theta: B_1[\vec{x}:=\vec{x}'] \Goes{} \theta')$ and
$(\theta': C\theta', \alpha \Goes{} \rho)$,
where $\rho$ is either some $\theta''$ or $diverge$,
then we have the following original computation:
\begin{center}
\begin{tabular}{ccc}
$\theta: B_1[\vec{x}:=\vec{x}'] \Goes{} \theta'$ \\
\cline{1-1}
$\theta: (B_1 \Or B_2)[\vec{x}:=\vec{x}'] \Goes{} \theta'$
  & \Sep &
  $\theta': C\theta', \alpha \Goes{} \rho$ \\
\cline{1-3}
\multicolumn{3}{c}{
  $\theta: if[\vec{x}((B_1 \Or B_2), C), \alpha \Goes{} \rho$
} \\
\end{tabular}
\end{center}
The corresponding computation with the transformed formula is:
\begin{center}
\begin{tabular}{ccc}
$\theta: B_1[\vec{x}:=\vec{x}'] \Goes{} \theta'$
  & \Sep \Sep &
  $\theta': C\theta', \alpha \Goes{} \rho$ \\
\cline{1-3}
\multicolumn{3}{c}{
  $\theta: if[\vec{x}](B_1, C) \alpha \Goes{} \rho$
} \\
\cline{1-3}
\multicolumn{3}{c}{
  $\theta: if[\vec{x}](B_1, C)
     \Or
     (\Not(\exists \vec{x} B_1)
      \And if[\vec{x}](B_2, C)), \alpha \Goes{} \rho$
} \\
\end{tabular}
\end{center}

If $(\theta: B_1[\vec{x}:=\vec{x}'] \Goes{} \theta')$ but
$(\theta': C\theta', \alpha \Goes{} fail)$,
then we have the following original computation:
\begin{center}
\begin{tabular}{ccc}
$\theta: B_1[\vec{x}:=\vec{x}'] \Goes{} \theta'$ \\
\cline{1-1}
$\theta: (B_1 \Or B_2)[\vec{x}:=\vec{x}'] \Goes{} \theta'$
  & \Sep &
  $\theta': C\theta', \alpha \Goes{} fail$ \\
\cline{1-3}
\multicolumn{3}{c}{
  $\theta: if[\vec{x}((B_1 \Or B_2), C), \alpha \Goes{} fail$
} \\
\end{tabular}
\end{center}
The corresponding computation with the transformed formula is:
\begin{center}
\begin{tabular}{ccc}
& &
  $\underline{\theta: B_1[\vec{x}:=\vec{x}'] \Goes{} \theta'}$ \\
& &
  $\vdots$ \\
& &
  $\overline{\theta: \exists \vec{x} B_1 \Goes{} \theta'}$ \\
& &
  $\overline{\theta:
     \Not(\exists \vec{x} B_1),
      if[\vec{x}](B_2, C), \alpha \Goes{} fail}$ \\
\cline{1-1} \cline{3-3}
$\theta: if[\vec{x}](B_1, C) \alpha \Goes{} fail$
  & \Sep &
  $\theta:
     (\Not(\exists \vec{x} B_1)
      \And if[\vec{x}](B_2, C)), \alpha \Goes{} fail$ \\
\cline{1-3}
\multicolumn{3}{c}{
  $\theta: if[\vec{x}](B_1, C)
     \Or
     (\Not(\exists \vec{x} B_1)
      \And if[\vec{x}](B_2, C)), \alpha \Goes{} fail$
} \\
\end{tabular}
\end{center}
where the computation of the left-hand premise of the bottommost
judgement is as follows:
\begin{center}
\begin{tabular}{ccc}
  $\theta: B_1[\vec{x}:=\vec{x}'] \Goes{} \theta'$ & \Sep &
  $\theta': C\theta, \alpha \Goes{} fail$ \\
\cline{1-3}
\multicolumn{3}{c}{
  $\theta: if[\vec{x}](B_1, C) \alpha \Goes{} fail$
} \\
\end{tabular}
\end{center}

We have very similar cases when 
$(\theta: B_1[\vec{x}:=\vec{x}'] \Goes{} fail)$ but
$(\theta: B_2[\vec{x}:=\vec{x}'] \Goes{} \theta')$,
depending on the result of $(\theta': C\theta', \alpha)$.

When both
$(\theta: B_1[\vec{x}:=\vec{x}'] \Goes{} fail)$ and
$(\theta: B_2[\vec{x}:=\vec{x}'] \Goes{} fail)$, we have
the following original computation:
\begin{center}
\begin{tabular}{c}
$\theta: B_1[\vec{x}:=\vec{x}'] \Goes{} fail$ \Sep
  $\theta: B_2[\vec{x}:=\vec{x}'] \Goes{} fail$ \\
\cline{1-1}
$\theta: (B_1 \Or B_2)[\vec{x}:=\vec{x}'] \Goes{} fail$ \\
$\overline{\theta: if[\vec{x}((B_1 \Or B_2), C), \alpha \Goes{} fail}$ \\
\end{tabular}
\end{center}
The corresponding computation with the transformed formula is:
\begin{center}
\begin{tabular}{ccc}
$\theta: B_1[\vec{x}:=\vec{x}'] \Goes{} fail$ \\
\cline{1-1}
$\theta: if[\vec{x}](B_1, C), \alpha \Goes{} fail$
  & \Sep &
  $\theta: (\Not(\exists \vec{x} B_1)
      \And if[\vec{x}](B_2, C)), \alpha \Goes{} fail$ \\
\cline{1-3}
\multicolumn{3}{c}{
  $\theta: if[\vec{x}](B_1, C)
     \Or
     (\Not(\exists \vec{x} B_1)
      \And if[\vec{x}](B_2, C)), \alpha \Goes{} fail$
} \\
\end{tabular}
\end{center}
where the computation of the right-hand premise of the
bottommost judgement is:
\begin{center}
\begin{tabular}{ccc}
$\theta: B_1[\vec{x}:=\vec{x}'] \Goes{} fail$ \\
\cline{1-1}
$\vdots$ & &
  $\theta: B_2[\vec{x}:=\vec{x}'] \Goes{} fail$ \\
\cline{1-1}
\cline{3-3}
$\theta: \exists \vec{x} B_1 \Goes{} fail$
  & \Sep &
  $\theta: if[\vec{x}](B_2, C), \alpha \Goes{} fail$ \\
\cline{1-3}
\multicolumn{3}{c}{
  $\theta: \Not(\exists \vec{x} B_1),
       if[\vec{x}](B_2, C), \alpha \Goes{} fail$
} \\
\multicolumn{3}{c}{
  $\overline{\theta: (\Not(\exists \vec{x} B_1)
      \And if[\vec{x}](B_2, C)), \alpha \Goes{} fail}$
} \\
\end{tabular}
\end{center}

The subcases in which a result of $diverge$ arises are similar
to those in which a result of $fail$ arises.

Case R5:  The two computations are
$(\theta: if[\vec{x}](B, C), \alpha \Goes{} \rho)$
and $(\theta: \exists \vec{x} (B \And C), \alpha \Goes{} \rho)$;
we must show that one implies the other.
We also know that $B$ is negated-disjunction.
There are two subcases.

If $(\theta: B[\vec{x}:=\vec{x}'] \Goes{} \theta')$,
then we have the following original computation:
\begin{center}
\begin{tabular}{ccc}
$\theta: B[\vec{x}:=\vec{x}'] \Goes{} \theta'$
  & \Sep &
  $\theta': C[\vec{x}:=\vec{x}']\theta', \alpha \Goes{} \rho$ \\
\cline{1-3}
\multicolumn{3}{c}{
  $\theta: if[\vec{x}](B, C), \alpha \Goes{} \rho$
  } \\
\end{tabular}
\end{center}
However, because $B$ is negated-disjunction, every computation
in the pessimistic semantics with
substitution and goal stack $(\theta: B[\vec{x}:=\vec{x}'], \alpha')$
must contain a substitution and goal stack
$(\theta': \alpha')$.  (See Lemma 4.6 of 
\cite{andrews-lnaf-tcs}.)  Thus we have
the following computation with the transformed formula:
\begin{center}
\begin{tabular}{c}
$\underline{\theta': C[\vec{x}:=\vec{x}']\theta', \alpha\theta' \Goes{} \rho}$ \\
$\vdots$ \\
\cline{1-1}
$\theta: B[\vec{x}:=\vec{x}'], C[\vec{x}:=\vec{x}'], \alpha \Goes{} \rho$ \\
\cline{1-1}
$\underline{\theta: (B \And C)[\vec{x}:=\vec{x}'], \alpha \Goes{} \rho}$ \\
$\vdots$ \\
$\overline{\theta: \exists \vec{x} (B \And C), \alpha \Goes{} \rho}$ \\
\end{tabular}
\end{center}
However, since $\theta'$ applies only to the free variables
of $B[\vec{x}:=\vec{x}']$, which are $\vec{x}'$, and $\alpha$
does not contain these variables, the topmost judgement is
equivalent to
$(\theta': C[\vec{x}:=\vec{x}']\theta', \alpha \Goes{} \rho)$.

Otherwise, $(\theta: B[\vec{x}:=\vec{x}'] \Goes{} \rho)$ where
$\rho$ is $fail$ or $diverge$.  In this subcase, the
bottom of the original computation is as follows:
\begin{center}
\begin{tabular}{c}
$\theta: B[\vec{x}:=\vec{x}'] \Goes{} \rho$ \\
\cline{1-1}
$\theta: if[\vec{x}](B, C), \alpha \Goes{} \rho$ \\
\end{tabular}
\end{center}
The bottom of the computation with the transformed formula is as follows:
\begin{center}
\begin{tabular}{c}
$\theta: B[\vec{x}:=\vec{x}'], C[\vec{x}:=\vec{x}'], \alpha \Goes{} \rho$ \\
\cline{1-1}
$\underline{\theta: (B \And C)[\vec{x}:=\vec{x}'], \alpha \Goes{} \rho}$ \\
$\vdots$ \\
$\overline{\theta: \exists \vec{x} (B \And C), \alpha \Goes{} \rho}$ \\
\end{tabular}
\end{center}
The presence of the extra formulas
($C[\vec{x}:=\vec{x}']$ and $\alpha$)
has no effect on the computation.
\end{proof}

\label{lastpage}
\end{document}